\documentclass[pre,twocolumn,superscriptaddress]{revtex4-1}
\usepackage{amsmath,amssymb,dsfont,graphicx,amssymb,mathrsfs,microtype,nicefrac}
\usepackage[svgnames,dvipsnames,x11names]{xcolor}
\usepackage[caption=false]{subfig}
\usepackage{grffile,floatrow,changepage,scrextend,comment,soul}
\usepackage{fancyhdr,tikz,wrapfig}
\usepackage{tabularx}
\usepackage[paperwidth=210mm,paperheight=297mm,centering,hmargin=2cm,vmargin=2.0cm]{geometry}

\usepackage[colorinlistoftodos]{todonotes}
\usepackage[colorlinks=true, allcolors=blue]{hyperref}

\newcommand{\add}[1]{\textcolor{black}{#1}}

\DeclareMathAlphabet{\mathpzc}{OT1}{pzc}{m}{it}

\newcommand{\norm}[1]{\left\lVert#1\right\rVert}

\colorlet{bluvi}{black}

\begin{document}

\title{Network Geometry}

\author{Mari\'an Bogu\~n\'a}
\affiliation{Departament de F\'isica de la Mat\`eria Condensada, Universitat de Barcelona, Mart\'i i Franqu\`es 1, E-08028 Barcelona, Spain}
\affiliation{Universitat de Barcelona Institute of Complex Systems (UBICS), Universitat de Barcelona, Barcelona, Spain}

\author{Ivan Bonamassa}
\affiliation{Department of Physics, Bar-Ilan University, 52900 Ramat-Gan, Israel}

\author{Manlio De Domenico}
\email[Corresponding author:~]{mdedomenico@fbk.eu}
\affiliation{CoMuNe Lab, Fondazione Bruno Kessler, Via Sommarive 18, 38123 Povo (TN), Italy}

\author{Shlomo Havlin}
\affiliation{Department of Physics, Bar-Ilan University, 52900 Ramat-Gan, Israel}

\author{Dmitri Krioukov}
\affiliation{Network Science Institute, Northeastern University, 177 Huntington avenue, Boston, MA, 022115} 
\affiliation{Department of Physics, Department of Mathematics, Department of Electrical~\&~Computer Engineering, Northeastern University, 110 Forsyth Street, 111 Dana Research Center, Boston, MA 02115, USA}

\author{M. {\'A}ngeles Serrano}
\affiliation{Departament de F\'isica de la Mat\`eria Condensada, Universitat de Barcelona, Mart\'i i Franqu\`es 1, E-08028 Barcelona, Spain}
\affiliation{Universitat de Barcelona Institute of Complex Systems (UBICS), Universitat de Barcelona, Barcelona, Spain}
\affiliation{ICREA, Passeig Llu\'is Companys 23, E-08010 Barcelona, Spain}

\date{\today}
\begin{abstract}
Real networks are finite metric spaces. Yet the geometry induced by shortest path distances in a network is definitely not its only geometry. Other forms of network geometry are the geometry of latent spaces underlying many networks, and the effective geometry induced by dynamical processes in networks. These three {\color{black}approaches to} network geometry are all intimately related, and all three of them have been found to be exceptionally efficient in discovering fractality, scale-invariance, self-similarity, and other forms of fundamental symmetries in networks. Network geometry is also of great utility in a variety of practical applications, ranging from the understanding how the brain works, to routing in the Internet. Here, we review the most important theoretical and practical developments dealing with these {\color{black}approaches to} network geometry in the last two decades, and offer perspectives on future research directions and challenges in this novel frontier in the study of complexity.
\end{abstract}

\maketitle

Many existing analytical and computational tools for the analysis of complex networks emerged from classical methods in statistical physics~\cite{cimini2019statistical}. 
Over the last two decades, these tools have proven essential for constructing models capable of reproducing the structural properties observed in many real-world networks~\cite{watts1998collective,barabasi1999emergence,ravasz2002hierarchical}, and for quantifying the importance of these properties for collective and critical phenomena in networks~\cite{dorogovtsev2008critical,gao2012networks,d2015anomalous,gao2016universal}. Many other complementary approaches have been also employed to study complex networks from different perspectives~\cite{bianconi2015interdisciplinary,estrada2012structure,garlaschelli2007organized,garlaschelli2009generalized,kalinin2018organized} leading to novel fundamental insights.
One such approach is {\em geometry}, the focus of this review.\\
\indent
The first evidence that complex networks possess some nontrivial geometric properties appeared with the discovery of their {\em self}-{\em similarity} under suitable scale transformations~\cite{song2005self}. 
Initially, {\em fractal geometry} was a major reservoir of methods and ideas. 
Besides boosting the study of transport phenomena in complex media~\cite{Gallos2007,condamin2007first}, the fractal geometric paradigm led to the definition of a 
reversible graph-theoretical renormalization procedure that helped researchers to classify networks into universality classes~\cite{radicchi2008complex,rozenfeld2010small}, as well as to better understand the growth mechanisms underlying their temporal evolution~\cite{song2006origins}.\\
\indent
Following the lines of this initial success, it was later found~\cite{serrano2008similarity} that network self-similarity can be explained at a more fundamental level in terms of {\em latent hyperbolic geometry}~\cite{krioukov2010hyperbolic}. 
This hidden metric space approach was successful in explaining, within a unified framework, the most common structural properties of many real networks~\cite{serrano2008similarity,krioukov2010hyperbolic}, their navigability~\cite{boguna2009navigability,gulyas2015navigable,allard2018,muscoloni2019navigability}, and their community~\cite{zuev2015emergence,garcia-perez2016hidden,wang2016hyperbolic,garcia-perez2018soft,alessandro2018leveraging} and multiscale~\cite{GarciaPerez2018} structures. 
Since the group of symmetries of hyperbolic spaces is isomorphic to the Lorentz group, the latent hyperbolicity of networks was advocated to explain not only their structural self-similarity, but also the dynamical laws of their growth~\cite{papadopoulos2012popularity,zuev2016hamiltonian,zheng2019b}, establishing certain duality relations~\cite{krioukov2013duality} reminiscent of the AdS/CFT correspondence.\\
\indent
In light of these advances, it is not surprising that the geometric approach led to many useful practical applications and novel theoretical insights. 
In the context of information or epidemic spreading, for example, the adoption of transport-based metrics and of the corresponding {\em diffusion geometries}~\cite{coifman2005geometric} is helping unfolding the spatiotemporal evolution of network-driven dynamical processes~\cite{brockmann2013hidden,taylor2015topological}, opening new research directions~\cite{dedomenico2017diffusion,hens2019spatiotemporal} in many neighboring areas of science.\\
\indent 
Contextual to these lines of research, there have been many other rapidly evolving research areas related to network geometry. The overall result of these developments is unprecedented cross-fertilization among many diverse fields of the natural sciences.
This rapid interdisciplinary progress suggests that now is about a right time to ground a milestone in network geometry research, from where to ponder on future challenges.

\begin{figure*}
\tikzstyle{mybox} = [draw=gray, fill=white, very thick,
    rectangle, rounded corners, inner sep=5pt, inner ysep=10pt]
\tikzstyle{fancytitle} =[fill=cyan!25, text=black]
	\begin{tikzpicture}
		\node [mybox] (box){
    			\begin{minipage}{0.98\textwidth}
    	\begin{tikzpicture}
    	\node [mybox] (box) at (0.45,0){
    	\includegraphics[width=0.282\linewidth]{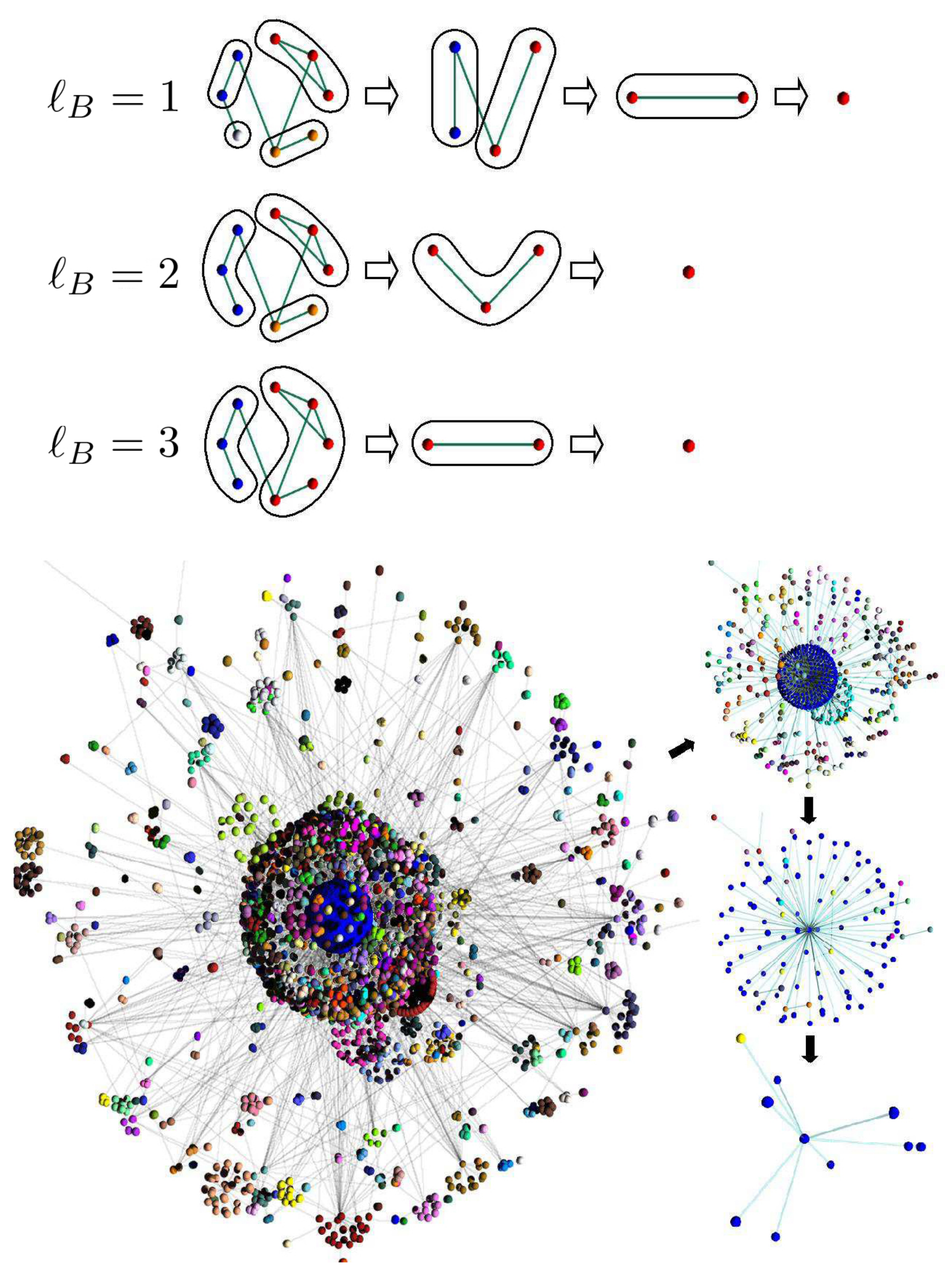}};
			\node[fancytitle,fill=Turquoise!30,right=9pt,draw,rounded corners,inner sep=2.5pt] at (+1.83,+3.22) {{\footnotesize (\textsc{A})}};
			\node[fancytitle,fill=none] at (-1.8,3.26) {\scriptsize{\textbf{a})}};
			\node[fancytitle,fill=none] at (-1.8,0.7) {\scriptsize{\textbf{b})}};
		\node [mybox] (box) at (5.97,0){
		\includegraphics[width=0.30\linewidth]{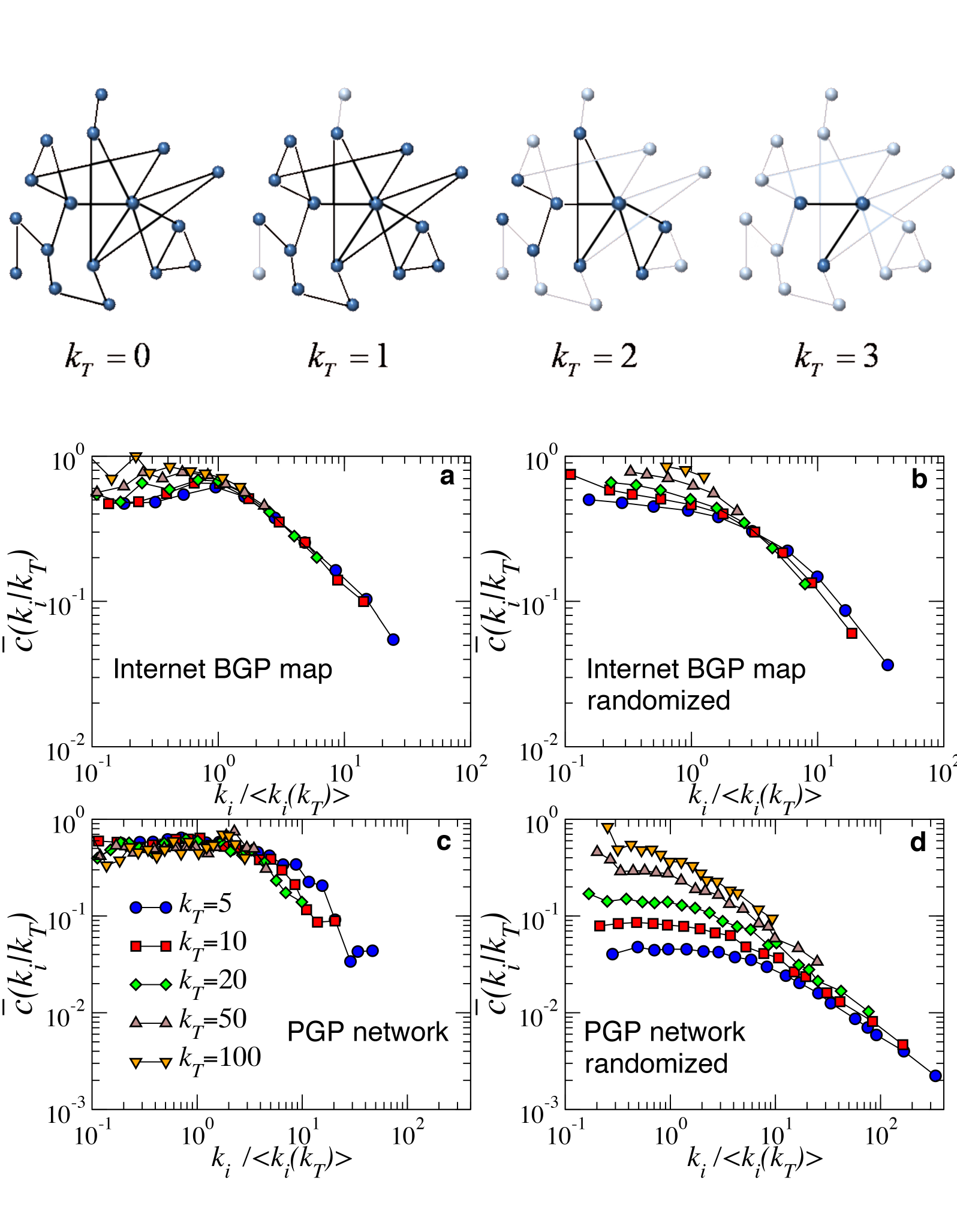}};
			\node[fancytitle,fill=Turquoise!30,right=9pt,draw,rounded corners,inner sep=2.5pt] at (7.45,+3.22) {{\footnotesize (\textsc{B})}};
			\node[fancytitle,fill=none] at (3.49,3.26) {\scriptsize{\textbf{a})}};
			\node[fancytitle,fill=none] at (3.49,0.7) {\scriptsize{\textbf{b})}};
		\node [mybox] (box) at (11.5,0){
    	\includegraphics[width=0.288\linewidth]{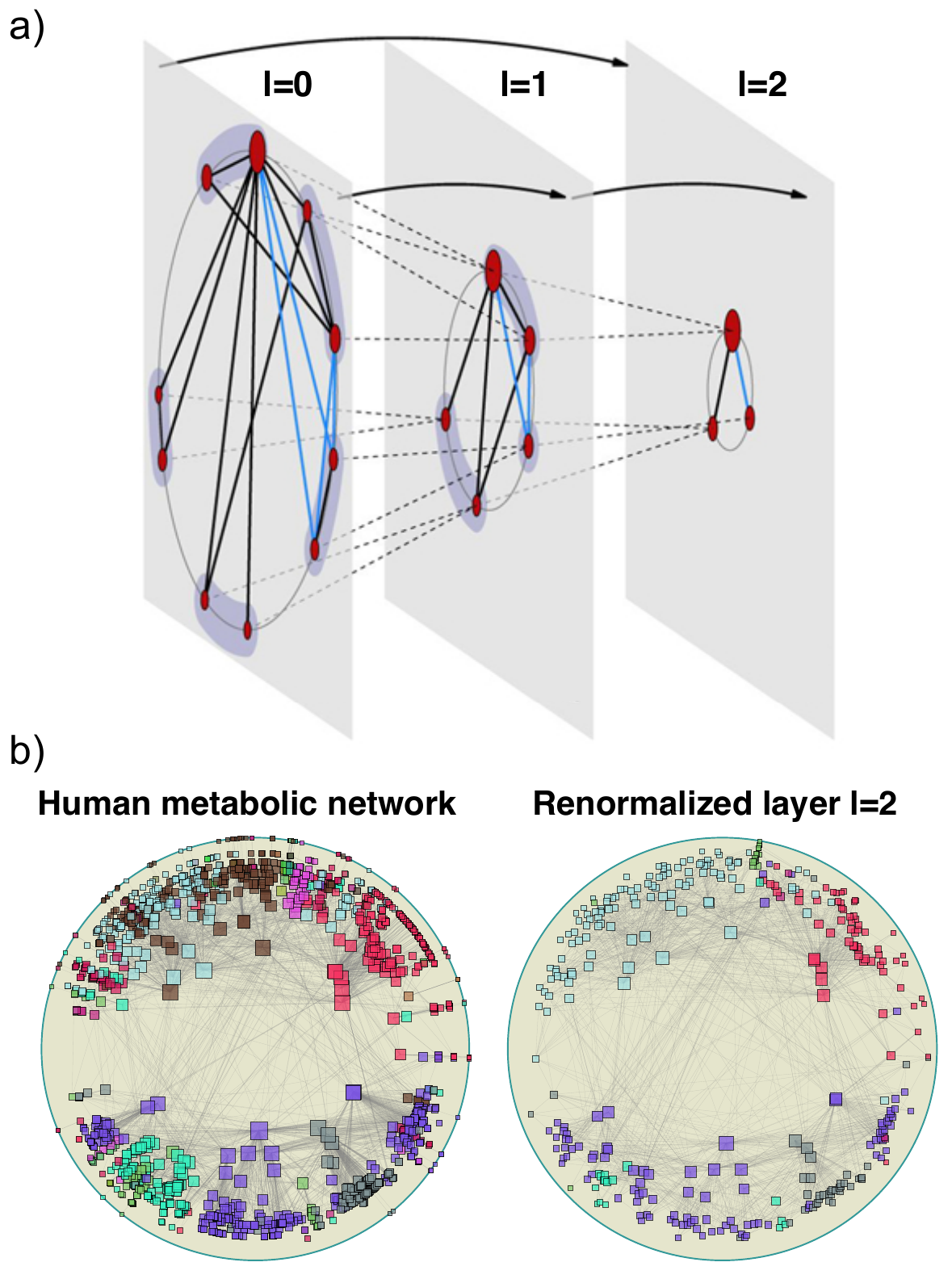}
    	\hspace*{-0.15cm}};
			\node[fancytitle,fill=Turquoise!30,right=9pt,draw,rounded corners,inner sep=2.5pt] at (13.0,+3.22) {{\footnotesize (\textsc{C})}};
	\end{tikzpicture} \\ \vspace{-0.25cm}
	\begin{center}\begin{addmargin}[0.85em]{0.85em}
	\footnotesize{
	(\textsc{A}) \textbf{Shortest-path-distance renormalization}. 
	\textbf{a}) Demonstration of the box covering technique (reproduced with permission from Ref.~\cite{song2005self}). The original network (first column) is tiled with the minimum number of boxes of diameter $\ell_B$ -- using e.g.\,the Maximum Excluded Mass Burning (MEMB) algorithm~\cite{song2007calculate} -- which are contracted into supernodes (second column) and then connected if at least one link existed between the ``parent'' tiles. 
	This coarse graining is repeatedly applied -- though for a limited number of steps due to small-worldness -- until the network is reduced to a single ``ancestral'' node. 
	\textbf{b}). Demonstration of the shortest-path-distance RG with $\ell_B=2$ applied to the entire WWW. 
	The network's structure remains statistically invariant under the renormalization (see Fig.~\ref{fig:1}). 
	(\textsc{B}) \textbf{Degree}-{\bf thresholding renormalization}. \textbf{a}) Subgraphs are obtained by removing all nodes with degrees below a given threshold $k_T$. This defines a hierarchy of nested subgraphs that are found to be self-similar in real complex networks. \textbf{b}) Data collapse of the clustering spectrum of the different subgraphs for the Internet and Border Gateway Protocol graphs (left column) and their randomized versions preserving the degree sequence (right column), for which the collapse is destroyed. The nice collapse of the clustering spectrum for real complex networks finds a natural explanation in their underlying geometry~\cite{serrano2008similarity}.
	(\textsc{C}) \textbf{Geometric renormalization}. \textbf{a}) Similarly to \textsc{A}), the renormalization transformation zooms out by changing the minimum length scale from that of the original network to a larger value, this time in the similarity space~\cite{GarciaPerez2018}. First, non-overlapping blocks of consecutive nodes are defined along the similarity circle. Second, the blocks are coarse-grained into supernodes. Each supernode is then placed within the angular region defined by the corresponding block so that the order of nodes is preserved. Finally, two supernodes are connected if any of their constituents were in the precursor layer. 
	\textbf{b}) Hyperbolic embedding of the human metabolic network and its renormalized layer $l=2$~\cite{GarciaPerez2018}. The colours of the nodes correspond to the community structure detected by the Louvain algorithm. Notice how the renormalized network preserves the original community structure despite being four times smaller.}
\end{addmargin}\end{center}
    \end{minipage}};
\node[fancytitle, right=10pt,draw,rounded corners,fill=ProcessBlue!35] at (box.north west) {\textsc{Box: Zooming out on complex networks}};
\end{tikzpicture}
\end{figure*} 
\indent
Here, we review three major research directions in network geometry: the self-similar fractal geometry of network structure (Sec.~\ref{sec:fractal}), the hyperbolic geometry of networks' latent spaces (Sec.~\ref{sec:hyperbolic}), and the geometry induced by dynamic processes, such as diffusion, in networks (Sec.~\ref{sec:dynamics}). Distances are all different in the three geometries, yet intimately related. They are, respectively, the \emph{shortest-path distances}, i.e.\ the hop lengths of shortest paths in a network, the \emph{latent distances}, i.e.\ the distances between network nodes in a latent space, and the \add{\emph{kinematic distances}, e.g.\ emerging from some spreading dynamics in a network such as reaction-diffusion or random search processes}.
We conclude the review with a discussion in Sec.~\ref{sec:discussion} of current advances in network geometry in a broader perspective, 
focusing on the most interesting open challenges, both theoretical and those that arise in applications, with the aim of emphasizing their rising impacts in physics and other fields of science.\\
\indent
Understood broadly, network geometry encompasses a great variety of diverse research directions, many of which are as important as the ones covered here, thus deserving separate reviews. Therefore, in a short review like this one, most of these directions must unfortunately be omitted. The most notable omissions include spatial networks, quantum gravity, graph curvature, geometrogenesis, graph embedding, topological data analysis, topological graph theory, and Gromov's $\delta$-hyperbolicity. However, at the end of Sec.~\ref{sec:discussion} we briefly comment on some of these topics that we find particularly important, promising, or intriguing. 

\section{Fractal geometry of network structure}\label{sec:fractal}
Self-similarity is a characteristic of certain geometric objects, known as fractals~\cite{feder2013fractals}, expressing the invariance of their forms under rescaling.
From snowflakes and ferns to turbulent flows~\cite{frisch1995turbulence} or critical phenomena near phase transitions~\cite{cardy1996scaling,lesne2011scale}, this {\em scale}-{\em free} property is ubiquitous among many natural systems, whose scaling factors are typically defined by the distance of the metric space in which they are naturally embedded. 

\begin{figure*}[t]
\tikzstyle{mybox} = [draw=lightgray, fill=white, very thick,
    rectangle, rounded corners, inner sep=1pt, inner ysep=10pt]
\tikzstyle{fancytitle} =[fill=cyan!25, text=black]
\begin{tikzpicture}
\node [mybox] (box){
\includegraphics[width=0.618\textwidth]{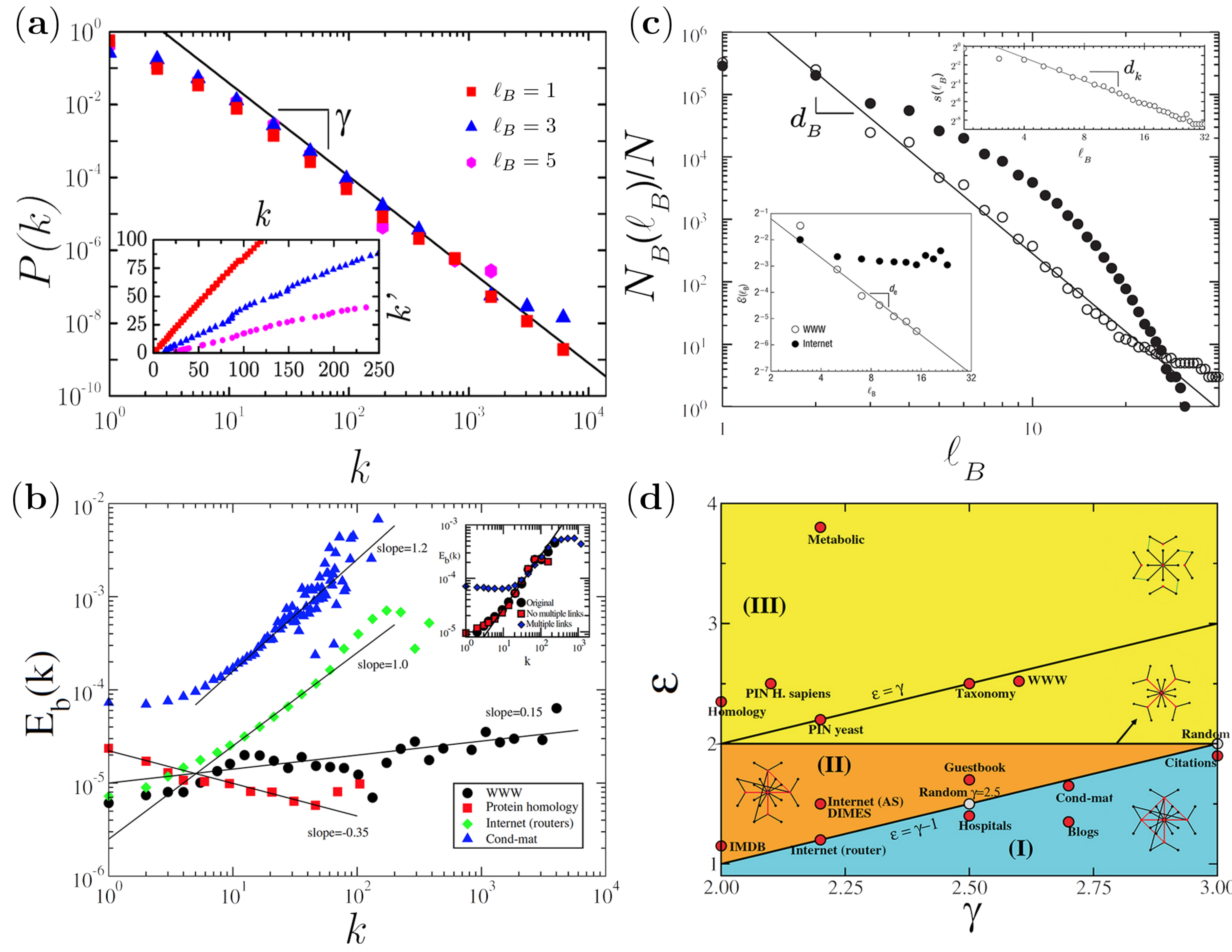}};
\node[fancytitle, right=9pt,draw,rounded corners] at (box.north west) {{\footnotesize \textsc{(A)} Structural self-similarity}};
\node [mybox] (box) at (8.20,0){
\includegraphics[width=0.337\textwidth]{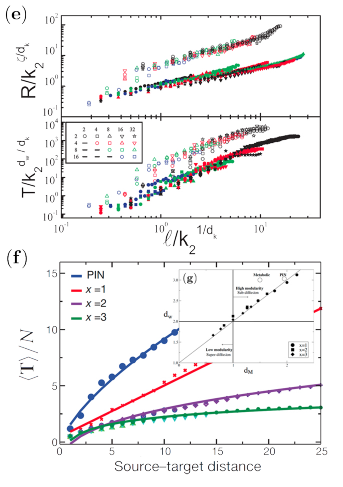}};
\node[fancytitle, right=9pt,draw,rounded corners] at (box.north west) {{\footnotesize \textsc{(B)} Transport theory}};
\end{tikzpicture}
\caption{\footnotesize \textsc{A})~\textbf{Structural self}-{\bf similarity}.
\textbf{a})~RG invariance of the WWW degree distribution and of its degree sequence (inset) for different box diameters $\ell_B$ (courtesy of Ref.~\cite{song2005self}). 
\textbf{b})~Degree-degree correlations measured in terms of $E(k)\sim k^{-(\epsilon-\gamma)}$ with $\epsilon>1$, obtained by integrating the degree-degree distribution $P(k,q)$ over $q${~\cite{rozenfeld2009fractal}}. 
(Inset) RG invariance of $E(k)$ for the Internet (router level) with $\ell_B=1$ (adapted with permission from Ref.~\cite{gallos2008scaling}). 
\textbf{c})~RG-scaling factors of the boxes' sizes, their degree sequence $s(\ell_B)\equiv k'/k$ (upper inset) and the hub-hub correlations $\mathcal{E}(\ell_B)$ (lower inset), versus the boxes diameter $\ell_B$ (with permission, Ref.~\cite{gallos2007review}).  
While fractal networks (e.g.\,the WWW, open circles) feature well-defined dimensions $d_B$, $d_k$ and $d_e$, non-fractal ones (e.g.\,Internet, filled circles) show an exponential (or faster) decay, i.e.\,$d_B,\,d_k\to\infty$ and $d_e\to0$ (courtesy of Ref.~\cite{song2006origins}). 
\textbf{d})~Classification of self-similar networks in the $(\gamma,\epsilon)$-plane (adapted with permission from Ref.~\cite{gallos2008scaling}). The lines $\epsilon_{rand}=\gamma-1$ and $\epsilon_{hier}=\gamma$ confine respectively the region I of random models and the area of hierarchical SF graphs; the scaling identity $\epsilon=2+d_e/d_k$ distinguishes non-fractal ($\epsilon\leq2$, region II) from fractal networks ($\epsilon>2$, region III).
\textsc{B})~\textbf{Transport theory}. 
\textbf{e})~Data collapse of the resistance (top) and the diffusion time (bottom) according to Eqs.~\eqref{eq:2p1} for the yeast PIN (open symbols) and the SHM model with $e=1$ (filled symbols). 
Different markers correspond to different ratios $k_1/k_2$, and different colors denote different values of $k_1$ (with permission from Ref.~\cite{Gallos2007}). 
\textbf{f})~Comparison between Eq.~\eqref{eq:3p1} and simulated diffusion times (markers) in the PIN and SHM model with $m=3$, $e=1$ and $x=1,2,3$ corresponding to $\zeta=1,\ln(3/2)/\ln3,0$, respectively (with permission from Ref.~{\cite{condamin2007first}}). 
\textbf{g})~Modularity vs transport in real-world and SHM model networks in the $(d_w,d_M)$ plane (courtesy of Ref.~{\cite{Gallos2007}}).}\label{fig:1}
\end{figure*}

In the context of complex networks, a similar discovery was made, though referring to the absence of a characteristic number of links $k$ per node rather than a length scale, as described by a fat-tailed degree distribution of the form $P(k)\sim k^{-\gamma}$ as $k\gg1$, with $\gamma\in(2,3)$. 
If, on one side, this scale-free (SF) property hinted at the existence of some degree of structural invariance under a suitable rescaling, the equally ubiquitous {\em small}-{\em world}~\cite{watts1998collective} property, i.e.~an average distance $\overline{\ell}$ between nodes growing logarithmically or slower~\cite{chung2002average,cohen2003scale,dorogovtsev2003metric} with the system's size $N$, 
contextually hindered this possibility---implying a diverging Hausdorff dimension---and led to the common belief that networks cannot be self-similar. 
Fortunately, this apparent contradiction proved to be a very prolific problem, whose analysis throughout tools of {\em fractal geometry}~\cite{mandelbrot1982fractal} has disclosed many fundamental insights in the hidden symmetries, renormalization and universality classes of complex networks.

\indent
\textbf{{\small\em Shortest-path-distance scaling and dimensions.}} In the graph theoretical sense, the number $\ell$ of edges along any shortest path connecting two nodes is a well-defined metric~\cite{harary1994graph}. 
It is known as the {\em shortest-path distance} or {\em chemical distance}~\cite{wiener1947structural,havlin1984cluster}, and it can be adopted to observe networks at different length scales. 
In this respect, in a seminal paper by Song {\em et al.}~\cite{song2005self}, it was shown that the process of zooming out on networks can be performed analogously to that of regular fractals by repeatedly coarse-graining the structure through optimal coverings of nodes made by non-overlapping boxes of diameter $\ell_B$ (Box, panel \textsc{A}). 
Under this renormalization group (RG) transformation~\cite{Wilson:1983,efrati2014real}, a surprisingly large variety of real-world and synthetic networks remain {\em statistically self}-{\em similar}, in the sense that their degree distribution~\cite{song2005self} and other mixing patterns~\cite{gallos2008scaling} are invariant (Fig.~\ref{fig:1}\textbf{a},\textbf{b}) over the available length scales.\\
\indent 
A primary consequence of this discovery was the characterization of networks featuring self-similarity through a set of {\em fractal dimensions}~\cite{bunde2012fractals} intimately related to their multi-scale organization and functioning. 
In what follows we review an essential selection of these network dimensions and of their significance. \\
\indent
We start from two crucial quantities, i.e.~the number of boxes (supernodes) $N'$ optimally covering a network and their degree sequence $\{k'_i\}_{i=1,\dots,N'}$, where each $k'$ corresponds to the degree of the hub contained in the ``parent'' box. 
Performing an RG step, $\mathcal{R}_B$, these quantities have power-law scaling
\begin{equation}\label{eq:1p1}
N\xrightarrow{\mathcal{R}_B}N'\sim \ell_B^{-d_B}N,\quad
k\xrightarrow{\mathcal{R}_B}k'\sim\ell_B^{-d_k}k,
\end{equation}
where the exponents $(d_B,\,d_k)$, called respectively the box and degree dimensions, characterize the (RG-invariant) SF degree distribution through the scaling identity $\gamma=1+d_B/d_k$~\cite{song2005self}. 
Besides unveiling a fundamental connection between SF and RG-invariant properties, the fractal analysis above yields a bird-eye view over the spectrum of different organization mechanisms underlying self-similar networks, by identifying two limiting cases~\cite{goh2006skeleton,kim2007fractality}: {\em i}) {\em fractal structures} when $d_B,\,d_k$ are both finite and non-zero (e.g.~biological systems, the WWW or social networks), and {\em ii}) {\em small}-{\em worlds} when $d_B,\,d_k\to\infty$ (e.g.~the Internet at the router level, or synthetic networks), implying an exponential decay (Fig.~\ref{fig:1}\textbf{c}) instead of Eq.~\eqref{eq:1p1}.\\
\indent
A more comprehensive picture is reached after analyzing structural fractality in terms of the different profiles of networks' degree-degree correlations~\cite{song2005self,yook2005self}. 
Fractal networks, in fact, feature a strong ``hub-hub repulsion'' which leads to disassortative structures, with hubs spread uniformly instead of being crumpled in a core as in pure small-worlds. 
In light of their RG-invariance~\cite{gallos2008scaling}, these correlation patterns can be bridle by a hub-hub dimension $d_e$ defined by the scaling factor $\mathcal{E}(\ell_B)\sim\ell_B^{-d_e}$ which accompanies the probability that two RG-boxes are connected through their hubs (Fig.~\ref{fig:1}\textbf{c}, lower inset). 
Scaling arguments analogous to those adopted for the triple $(d_B,\,d_k,\,\gamma)$ allow here to bridge the correlation exponent $\epsilon>1$---defined by the degree-degree distribution $P(k,q)\sim k^{-\gamma+1}q^{-\epsilon}$---to the hub-hub dimension through the identity $\epsilon=2+d_e/d_k$~\cite{gallos2008scaling}. 
These results allow to classify a large variety of networks according to RG-invariant properties related to their large-scale organization, as shown in the $(\gamma,\epsilon)$ phase diagram in Fig.~\ref{fig:1}\textbf{d}.\\
\indent
One last, yet essential feature of networks' fractal geometry is the existence of a larger group of self-similar symmetries leaving their meso-scale organization invariant. 
The RG transform, in fact, identifies a hierarchy of modular configurations into which networks can be optimally partitioned at increasing length scales. 
The scale-invariance of this tiling can be quantified via the power-law scaling $\mathcal{Q}(\ell_B)\sim\ell_B^{d_M}$~\cite{Gallos2007,galvao2010modularity}, where $d_M$ is called the {\em modular dimension} and $\mathcal{Q}(\ell_B)$ is a modularity factor~\cite{newman2006modularity,fortunato2010community} maximized by the box covering. 
In particular, the value $d_M=1$ (characterizing regular lattices) identifies the borderline case separating modular structures ($d_M>1$, typical of e.g.~biological network) from non-modular ones ($d_M<1$), signaling increasing levels of small-worldness for decreasing values of $d_M$. 
We will see in what follows that $d_M$, unlike the other ``structural'' dimensions, is directly connected to other dynamical exponents (Fig.~\ref{fig:1}\textbf{g}) characterizing transport in complex media.

\begin{figure*}
\tikzstyle{mybox} = [draw=lightgray, fill=white, very thick,
    rectangle, rounded corners, inner sep=1pt, inner ysep=10pt]
\tikzstyle{fancytitle} =[fill=cyan!25, text=black]
\begin{tikzpicture}
\node [mybox] (box){
\includegraphics[width=0.233\textwidth]{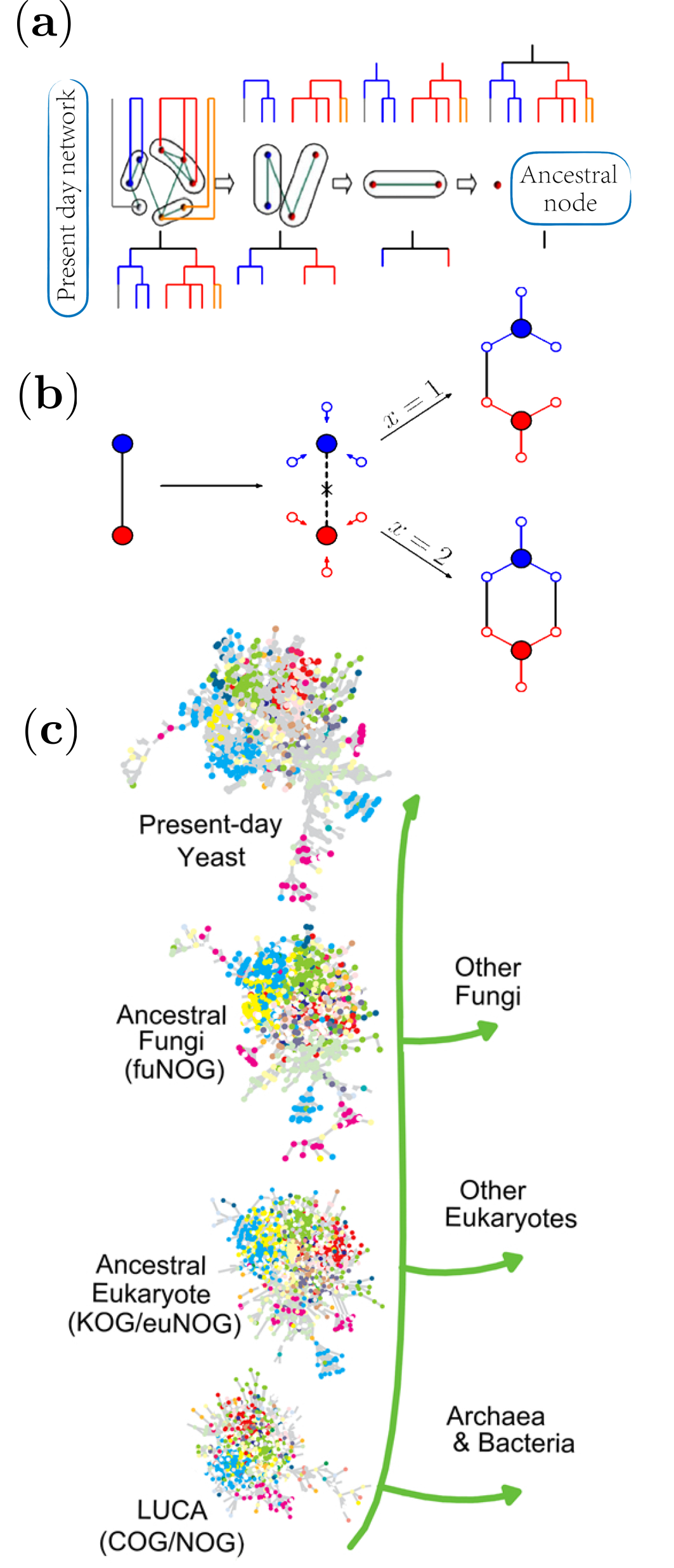}};
\node[fancytitle, right=9pt,draw,rounded corners] at (box.north west) {{\footnotesize \textsc{(A)} Evolution}};
\node [mybox] (box) at (8.45,0){
\includegraphics[width=0.754\textwidth]{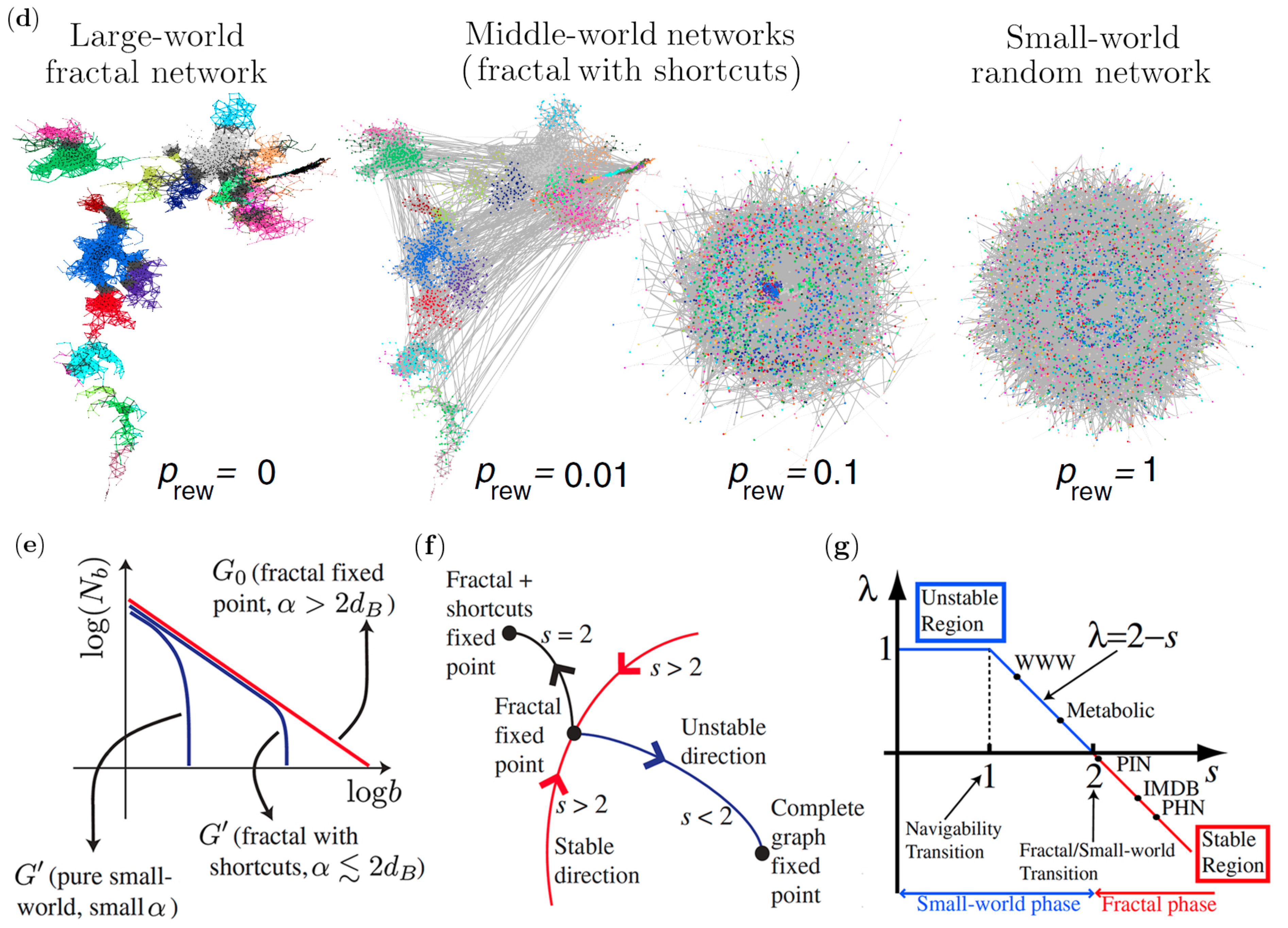}};
\node[fancytitle, right=9pt,draw,rounded corners] at (box.north west) {{\footnotesize \textsc{(B)} Shortest-path-distance RG flow \& Complex network universality classes}};
\end{tikzpicture}
\caption{\footnotesize 
\textsc{A})~\textbf{Evolution}. 
\textbf{a})~Demonstration of network evolution as inverse of the shortest-path-distance RG procedure (Box \textsc{A}); the arrow of time increases from right to left. 
\textbf{b})~Seeds of the SHM model for network growth~\cite{song2006origins}. At each step and for every link, a node produces $m$ offsprings. The original link is then removed with probability $e$, and $x$ new links between randomly selected nodes of the new generation are added. Illustration: $m=3$, $e=1$. While $e$ tunes between pure fractals ($e=1$) and pure small-worlds ($e=0$), $x$ rules the degree of modularity, so that $x=1$ yields tree-like structures and shortcuts among modules appear for $x>1$. 
\textbf{c})~Phylogenetic tree showing the evolutionary path of the {\em S. cerevisiae} PIN network, reconstructed by combining the SHM model with the RG process (adapted with permission from Ref.~{\cite{jin2013evolutionary}}). Nodes' colors represent different functional categories. 
\textsc{B})~\textbf{RG flow and universality}. 
\textbf{d})~Fractal to small-world transition, obtained by randomly rewiring a fraction $p_{rev}$ of the total number of links. While modularity is rapidly lost, small-worldness is rapidly gained, emphasizing the trade-off between these network phases (courtesy of Ref.~\cite{gallos2012small}).
\textbf{e})~Crossover from the power-law scaling in Eq.~\eqref{eq:1p1} to an exponential decay in fractal networks upon the random addition of shortcuts according to the probability $P(\ell)=A\ell^{-\alpha}$. 
\textbf{f})~RG-flow diagram in the space of configurations (reproduced with permission, Ref.~{\cite{rozenfeld2010small}}).
In the stable phase ($s\equiv\alpha/d_B>2$) the RG flows towards the fractal fixed point, while in the unstable phase ($s<2$) it flows towards a complete graph. 
For $s\simeq2$ the scaling has an exponential cut-off at a characteristic scale (reflected in \textbf{e})) indicating that the network is a global small-world and a fractal at small length scales.
\textbf{g})~{\em Phase diagram and universality}. 
The stability analysis of the RG-flow (enshrined in the stability exponent $\lambda$) leads to Eqs.~\eqref{eq:4p1}--\eqref{eq:5p1}, in terms of which a navigability ($s=1$) and a fractal to small-world ($s=2$) thresholds are identified, leading to a classification of real-world complex networks in the space of network configurations (courtesy of Ref.~\cite{rozenfeld2010small}). 
\label{fig:self2}}
\end{figure*}

\textbf{{\small\em Networks' functionality and evolution.}} 
If on the one hand the shortest-path-distance RG enabled to place on more fundamental grounds the networks' mixing patterns, on another one it raised several puzzles of interpretation~\cite{strogatz2005complex}. 
Besides questioning the functional significance encoded in the modules detected by the RG, it soon proved impossible to understand their emergence in terms of popular mechanisms of network growth~\cite{dorogovtsev2003evolution} like e.g.~the ``rich-get-richer'' principle of preferential attachment (PA), or the ``democratic'' wiring of Erd\H{o}s-R\'enyi (ER) networks, which are generally dominated by small-worldness.\\
\indent
The first issue was soon clarified by analyzing diverse biological networks~\cite{song2006origins,Gallos2007,galvao2010modularity}, where the RG transform has led to identify hierarchies of modules closely related to their known biochemical annotations, with a resolution as accurate as that of other clustering algorithms~\cite{ravasz2002hierarchical,newman2006modularity,fortunato2010community}. 
This is probably best manifested in Ref.~\cite{galvao2010modularity}, where the RG analysis has enabled researchers to develop the very first integrated multiscale-view of the network of human cell differentiation, raising the possibility of identifying hitherto unknown functional relations between previously unrelated cellular domains.\\ 
\indent
The problem of growth, instead, has been elegantly solved by rooting on the observation that modular networks are also fractal (though the converse is not necessarily true). 
Building on this property, Song, Havlin \& Makse (SHM) formulated~\cite{song2006origins,Gallos2007} a multiplicative process of network growth whose dynamics follow the inverse of the shortest-path-distance RG transform (Fig.~\ref{fig:self2}\textbf{a},\,\textbf{b}), where hubs acquire new connections by linking preferentially with less connected nodes under the noisy appearance (as exemplified in the Watts-Strogatz model) of randomly placed shortcuts. 
The net result is that hubs are deep buried into modules whose low-degree nodes are the inter-module connectors, resulting into SF networks which are fractal and modular up to a cut-off scale above which they become global small-worlds (Fig.~\ref{fig:self2}\textbf{e}). 
The SHM model represents a theoretical benchmark for understanding the self-similar patterns observed in real-world systems in terms of the microscopic growth rates controlling their dynamics~\cite{rozenfeld2009fractal}, and it has raised important evolutionary implications. 
Besides highlighting the evolutionary drive of many biological networks~\cite{song2006origins} towards fractal modular structures---which maximize their robustness against random failures~\cite{kitano2002systems}---the SHM model has led to uncover the inherently fractal geometric nature of the duplication-divergence mechanisms, suggesting that fractality and multiplicative growth are essential features of biological network evolution. 
A notable result in this respect has appeared in Ref.~{\cite{jin2013evolutionary}}, where the present day structures of different protein-protein interaction networks (PINs) have been successfully reconstructed (Fig.~\ref{fig:self2}\textbf{c}) starting from their primitive ancestors identified via the shortest-path-distance RG technique. 

\textbf{{\small\em Transport in self-similar media.}} 
Discovering the fractal geometry of networks has also had significant implications in the study of transport in complex media, a notoriously hard theoretical problem~\cite{villani2008optimal}. 
The existence of an underlying self-similar symmetry, in fact, allows to circumvent (at least at first glances) the search for exact solutions and to attack the problem by means of scaling arguments. 
A first result in this direction was presented in Ref.~\cite{Gallos2007}, where Gallos {\em et al.}~raised evidences about the scale-invariant forms featured by the mean-diffusion time, $T\sim\ell^{d_w}$, and the average resistance, $R\sim\ell^\zeta$, experienced by a blind ant traveling a shortest-path-distance $\ell$, where $d_w$ and $\zeta$ are respectively called the {\em walk} and {\em conductivity dimensions}~\cite{ben2000diffusion} of the underlying media. 
Combining these relations with Eqs.~\eqref{eq:1p1}, the observables $T$ and $R$ re-scale respectively as $T'/T\sim(N'/N)^{d_w/d_B}$ and $R/R'\sim(N'/N)^{\zeta/d_B}$ under RG transformations.
A direct evaluation of $d_w$ and $\zeta$ in biological and synthetic (SHM) networks with finite box dimension $d_B$, reveals~\cite{Gallos2007} that the triple $(d_B,d_w,\zeta)$ obeys the {\em Einstein identity}~\cite{haynes2009generalization} $\zeta=d_w-d_B$, elegantly relating static and dynamic properties of transport in complex media. 
Thanks to this insight, it proved possible to predict the dependence of $T$ and $R$ on microscopic network features---e.g.~nodes' degrees and chemical distance between them---leading to the scaling relations
\begin{equation}\label{eq:2p1}
\begin{aligned}
T(\ell;k_1,k_2)&\,= k_2^{d_w/d_k}\mathpzc{f}_T\big(\ell/k_2^{1/d_k}\big),\\
R(\ell;k_1,k_2)&\,= k_2^{\zeta/d_k}\mathpzc{f}_R\big(\ell/k_2^{1/d_k}\big),
\end{aligned}
\end{equation}
\noindent 
where $\mathpzc{f}_{T,R}$ are scaling functions, finding excellent agreement with the collapse of real--world networks data (Fig.~\ref{fig:1}\textbf{e}). 
A more complete scenario with respect to the above has appeared in Ref.~\cite{condamin2007first}, where a general theory of transport in complex media was developed based on the assumption of self-similarity, culminating in the exact, finite-size expression
\begin{equation}\label{eq:3p1}
T(\ell)\sim\begin{cases}
N\big(\mathpzc{a}+\mathrm{sgn}(\zeta)\mathpzc{b}\ell^{\zeta}\big),\quad\zeta\gtrless0,\\
\quad N\big(\mathpzc{a}+\mathpzc{b}\ln\ell\big),\qquad  \zeta=0,
\end{cases}
\end{equation}
\noindent 
where $\mathpzc{a},\,\mathpzc{b}$ are positive constants. 
Eq.~\eqref{eq:3p1} describes a universal scaling law for transport in complex media~\cite{masuda2017random} having a finite box-dimension (Fig.~\ref{fig:1}\textbf{f}).\\ 
\indent
Besides the theoretical importance, these results have offered new methods of fractal geometry to relate the kinetics of transport-limited processes in real-world systems to their interaction topologies. 
Eqs.~\eqref{eq:2p1}, in particular, allows bridging the degree of modularity of networks to their efficiency of transport~\cite{Gallos2007}. 
This is best epitomized by the identity $d_w=1+d_M$ (Fig.~\ref{fig:1}\textbf{g}) which shows how high levels of modularity ($d_M>1$) generally result into sub-diffusive dynamics ($d_w>2$). 
This simple result has had significant implications in e.g.~the characterization of the flux responses of metabolites~\cite{Gallos2007}, the community detection in global small-world social networks~\cite{gallos2013imdb}, and it has provided intriguing ideas for addressing the long-standing conundrum of the highly modular yet globally optimal organization of functional brain networks~\cite{gallos2012small,avena-koenigsberger2017communication}. We will return to this topic in Sec.~\ref{sec:hyperbolic} and later in the discussions, when addressing the problem of the geometric routing and organization of the human brain.

\indent 
\textbf{{\small\em RG-flow.}} The repeated application of the shortest-path-distance RG transformation, $\mathcal{R}_b$, identifies a flow in the space of graph configurations which, much like the case of critical phenomena~\cite{Wilson:1983,stanley1999scaling,efrati2014real}, enables a classification of network topologies into {\em universality classes}. 
Scale-invariance and self-similarity, in fact, are natural symmetries featured by the fixed points of the RG flow, whose stability against small perturbations enshrines universal features.\\
\indent 
A preliminary study aiming at exploring this direction was carried out by Radicchi {\em et al.} in Ref.~\cite{radicchi2008complex}. 
In this work, a generalization of finite-size scaling methods was introduced for analyzing the RG-flow beyond the limiting constraint of a few RG-steps due to the networks' small-worldness. 
Scrutinizing a large set of real-world and artificial networks~\cite{radicchi2009renormalization}, a coherent picture about their stability under the RG flow emerged after looking at their fluctuation ($\gamma'$) and correlation length ($\nu$) critical exponents~\cite{stanley1971introduction}. 
While $\nu=\gamma'=2$ in every small-world network (say, ER graphs, Watts-Strogatz networks, the PA model, etc.), different values of the couple $(\nu,\gamma')$ were shown to be featured by different fractal structures (e.g.\,the SHM model, Apollonian networks, percolating clusters, etc.).
For the SHM model, in particular, one can prove~\cite{radicchi2009renormalization} that $\nu=\gamma'=1$ for every SF exponent, while $\nu=\gamma'=2$ after any arbitrarily small random rewiring, suggesting that fractal topologies are indeed {\em unstable} fixed points of the shortest-path-distance RG flow. 
For increasing fractions of randomly drawn shortcuts, in fact, fractal networks rapidly crossover to more and more compact architectures (Fig.~\ref{fig:self2}\textbf{d}) having weaker and weaker modularity.
This leads to networks whose fractal scaling $\bar{\ell}\sim {N_0}^{1/d_B}$ is observed only up to a certain cut-off length-scale before a global small-world behavior $\bar{\ell}\sim \ln N_0$ is found, as depicted in Fig.~\ref{fig:self2}\textbf{e}.\\
\indent
These promising results have prepared the ground for the firm RG theory presented in Ref.~\cite{rozenfeld2010small}, which elegantly elucidates the universal features underlying the fractal to small-world transition in complex networks. 
In this work, Rozenfeld {\em et al.} have showed that adding shortcuts to a fractal network $G_0$ with a probability $\mathpzc{p}(\ell)=\mathpzc{A}\ell^{-\alpha}$, where $\mathpzc{A}$ is a normalization constant, brings the RG trajectories either to converge towards $G_0$ or to transform it into a complete graph---a trivially stable fixed point of the RG-flow---depending on the value of the exponent $\alpha>0$. 
To quantify these ideas, the authors focused on the renormalized distribution of shortcuts after one RG step which, in the formal limit of $\ell_B\to\infty$, leads to the fixed point equation
\begin{equation}\label{eq:4p1}
p^*(\ell)\equiv1-\lim_{x\to\infty} \mathrm{exp}\Big[-\mathpzc{C}(\ell)x^{2d_B/\alpha-1}\Big],
\end{equation}
\noindent 
where $\mathpzc{C}(\ell)\equiv \mathpzc{A}^{2d_B/\alpha}\ell^{-2d_B}$ and $x\equiv \mathpzc{A}^{-1}(\ell_B\ell)^{\alpha}$. 
Eq.~\eqref{eq:4p1} has three distinct solutions (Fig.~\ref{fig:self2}\textbf{f}), depending on the value of the parameter $s\equiv \alpha/d_B$: 
\begin{itemize}
    \item[$\bullet$] if $s>2$, then $p^*(\ell)=0$ and the RG-flow converges again towards the fractal network $G_0$;
    \item[$\bullet$] if $s<2$, then $p^*(\ell)=1$ and the RG-flow converges towards the complete graph fixed point; 
    \item[$\bullet$] if $s=2$, then the RG-flow has one non-trivial stable fixed point $G'$, consisting of $G_0$ dressed with shortcuts following $p^*(\ell)=1-\mathrm{exp}(-\mathpzc{A}\ell^{-2d_B})$.
\end{itemize}

\indent 
To gain further insights into these three network phases, the authors analyzed the flow of the difference $z_b-z_0$ between the average degrees in $G_0$ and in the renormalized network $G_b=\mathcal{R}_b(G')$. 
They found that $z_b-z_0=(z'-z_0)\mathcal{D}(x_b)$ where, in the thermodynamic limit, the function $\mathcal{D}(x_b)$ scales with the relative network size as $\mathcal{D}(x_b)\sim x_b^\lambda$ and the RG exponent $\lambda$ depends on the shortcut exponent $\alpha$ as
\begin{equation}\label{eq:5p1}
\lambda=\begin{cases}\begin{aligned}
&\,1,\qquad\,\,\, \text{if}\,\,\,s\leq1,\\
&\,2-s,\quad \!\text{if}\,\,\,s>1. 
\end{aligned}\end{cases}
\end{equation}
\noindent 
As summarized in Fig.~\ref{fig:self2}\textbf{g}, Eq.~\eqref{eq:5p1} identifies two transitions in the space of network configurations: $i$) a small-world to fractal transition at $s=2$, equivalently at $\alpha=2d_B$, separating the stable ($\lambda<0$, $s>2$) phase of compact topologies from the unstable phase ($\lambda>0$, $s<2$) of modular structures; $ii$) a {\em navigability transition} at $s=1$, equivalently at $\alpha=d_B$, identifying the network analogue of Kleinberg's optimal point~\cite{kleinberg2000navigation}.\\
\indent 
Besides raising theoretical questions regarding the characterization of these configurational transitions, the RG theory sketched above provides an indirect method for extracting information about the distribution of shortcuts in fractal real-world networks -- a crucial ingredient for understanding information flow and optimal search~\cite{viswanathan1999optimizing} -- and to determine their approximate location in the space of network configurations. 
The results of this analysis applied to the WWW, the metabolic network of {\em E.\,coli}, a yeast protein interaction network (PIN), the actors network of IMDB, and the protein homology network (PHN), are depicted in Fig.~\ref{fig:self2}\textbf{g}. 
The results show, in particular, that the WWW (Fig.~\ref{fig:1}\textbf{c}) is fractal up to a given length scale, but it is also sufficiently randomized for hosting an optimal flow, as manifested by its proximity in the $(\lambda,s)$ plane to the navigability threshold. 
As we will see in the next section, a profound discovery unveiling the latent geometricity of complex networks has further deepened our understanding about their navigability and optimal routing, providing unfamiliar perspectives in the study of networks' structure and function.

\section{Hyperbolic geometry of network latent spaces}\label{sec:hyperbolic}

The previous section deals with self-similarity and fractal properties of the structure of real-world networks. There exist deep connections between self-similarity and hyperbolic geometry that have been well explored in mathematics~\cite{gromov1987hyperbolic,nekrashevych2005similar,buyalo2007elements}.
One connection goes via the observation that any Gromov-hyperbolic space has a boundary at infinity which is always a self-similar metric space~\cite{buyalo2007elements}. 
Another connection is that self-similar groups can be always represented as the groups of automorphisms of trees, which are the simplest example of discrete hyperbolic spaces~\cite{nekrashevych2005similar}.
The connections between hyperbolicity and self-similar sets, fractals, and similar objects 
goes also through the idea that rescaling is (approximately) an isometry transformation in (coarse) hyperbolic geometry~\cite{grigorchuk2015similar}.\\ 
\indent
Yet hyperbolic geometry turned out to be not the geometry of the observable structure of real-world networks discussed in the previous section, but the geometry of their latent spaces that we discuss next. 
The two geometries turned out to be intimately related because network paths that follow hyperbolic geodesics in the latent space are shortest paths in the network with high probability~\cite{boguna2009navigability,boguna2009navigating,krioukov2010hyperbolic,boguna2010sustaining}. Whenever this happens, we say that the network is congruent with its underlying latent geometry.

\begin{figure*}[t]
\tikzstyle{mybox} = [draw=gray, fill=white, very thick,
    rectangle, rounded corners, inner sep=2pt, inner ysep=5pt]
\tikzstyle{fancytitle} =[fill=cyan!25, text=black]
	\begin{tikzpicture}
    		\node [mybox] (box){
    		\includegraphics[width=0.9\linewidth]{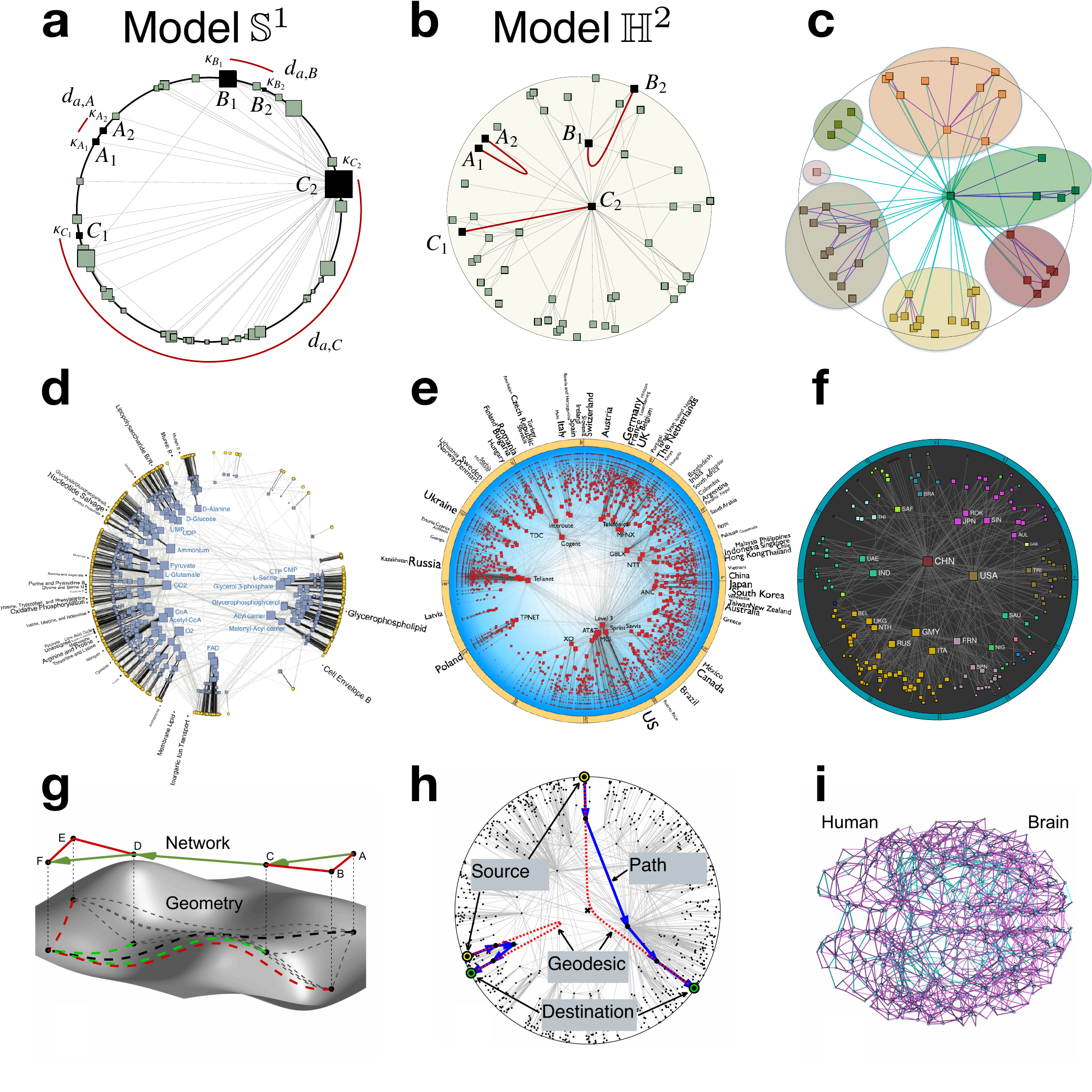}};
\end{tikzpicture}
\caption{\footnotesize 
\textbf{Networks in latent geometry.} \textbf{a--b) Models.} \textbf{a}) Model $\mathbb{S}^1$ (one-dimensional sphere). The $\mathbb{S}^1$ distances $d_a$ between pairs of nodes A1-A2, B1-B2 and C1-C2 are highlighted. The size of a node is proportional to its expected degree $\kappa$. \textbf{b}) Model~$\mathbb{H}^{2}$ (two-dimensional hyperbolic disk). All the shown pairs of nodes are at the same hyperbolic $\mathbb{H}^{2}$ distance, highlighted. Higher-degree nodes are positioned closer to the centre. The angular $\mathbb{S}^1$ distances between the corresponding node pairs are the same in~\textbf{a} and~\textbf{b}. \textbf{c)~Critical Gap Method (CGM).} Nodes are partitioned into different groups separated by void angular gaps. The modularity of the partition is computed by comparing the number of links within the communities (purple links) to the number of links between nodes in different communities (green links). The partition with the highest modularity is selected. \textbf{d--f) Embeddings of real networks and their community organization.}  \textbf{d}) Metabolic network of bacterium {\it E.~coli}~\cite{serrano2012uncovering}. Yellow circles represent reactions whereas blue  squares are metabolites. The name of a pathway is located at the average angular position of all the reactions belonging to it. \textbf{e}) Internet at the Autonomous Systems level~\cite{boguna2010sustaining}: The name of each country is located at the average angular position of its Autonomous Systems. \textbf{f}) World trade map in 2013~\cite{garcia-perez2016hidden}: Different node colors correspond to different communities detected by the CGM method.
\textbf{g--i) Geometric routing.} \textbf{g}) Finding a path from~$A$ to~$F$ in the network is done step-by-step. The source node~$A$ first checks which one of its neighbors in the network, $B$ or $C$, is closer to the destination node~$F$ in the underlying geometry, where the geodesics are shown by the dashed curves. Node~$C$ is closer, so that it is the next hop on the path from~$A$ to~$F$. Node~$C$ then performs similar calculations to find that~$D$ is closest to~$F$ among its network neighbors~$A,B,D$, so that $D$ is the next hop after~$C$, and also a penultimate hop because it is a neighbor of~$F$. \textbf{h)} Proximity of shortest paths in hyperbolic networks to hyperbolic geodesics. The blue arrows show the paths that geometric routing finds between a couple of source-destination pairs in a hyperbolic network. The found paths are also the shortest paths in the network in terms of the number of hops. The hyperbolic geodesics between the corresponding sources and destinations in the hyperbolic plane are shown as the dashed red curves. \textbf{i)} Navigation skeleton of the human brain~\cite{gulyas2015navigable}. The structural network of the human brain with links colored depending on whether they belong (magenta) or do not belong (cyan) to the minimal network that enables maximal navigability in the brain.}
\label{part2_Fig_comm}
\end{figure*}

\indent
\textbf{\em Models.}  Latent spaces have been employed for nearly a century to model homophily in social networks~\cite{sorokin1927social,mcfarland1973social,hoff2002latent}. In these models, nodes are positioned in a similarity space, while connections between them are random, but they are the more likely, the closer the two nodes in the space, i.e.\ more similar nodes are more likely to be connected.\\
\indent
These models are known as \emph{(soft) random geometric graphs} in mathematics, where they have been extensively explored~\cite{penrose2003random}. In the simplest random geometric graph model, $n$ nodes are placed uniformly at random on the interval $[0,n]$ with periodic boundary conditions, i.e.\ the circle~$\mathbb{S}^1$. The node pairs are then connected if the distance between them on the circle is less than a parameter $\mu>0$ controlling the average degree ($\langle k \rangle=2\mu$). The model yields networks with non-vanishing clustering ($\langle c \rangle = 3/4$) and average shortest path length that scales linearly with network size. These networks are thus large worlds.\\
\indent
From the statistical physics perspective, this simplest possible latent-space network model is the zero-temperature ($\beta\to\infty$) limit of a more general entropy-maximizing probabilistic mixture of grand canonical ensembles with the Fermi-Dirac probability of connection between nodes~$i$ and~$j$:
\begin{equation}\label{eq:conn_prob_FD}
p_{ij}=\frac{1}{e^{\beta(\varepsilon_{ij}-\mu)} + 1}.
\end{equation}
In this ensemble, edges are fermions with energies
\begin{equation}\label{eq:energy_rgg}
  \varepsilon_{ij}=f(x_{ij}),
\end{equation}
where $f(x_{ij})$ can be any function of distances $x_{ij}$ 
between nodes on the circle, $\beta$ is the inverse of the temperature fixing the average energy, and $\mu$ is the chemical potential controlling the expected number of particles-edges and thus the average degree.\\
\indent A choice of $f(x)$ defines network properties in the ensemble, and it was shown 
in~\cite{boguna2020small} that the {\it necessary and sufficient} conditions for networks in the model to be sparse small-worlds with non-vanishing clustering, are $f(x) \propto \ln{x}$ and $\beta \in(1,2)$. More precisely, if $f(x) \propto \ln{x}$, then clustering in the $n\to\infty$ limit is zero for $\beta\leq1$, but an increasing positive function of $\beta>1$, while networks are small-worlds whenever $\beta<2$.\\
\indent
The distribution of node degrees in such models is homogeneous, but it can be modified to yield \emph{any} degree distribution. This modification sets the edge energy given in Eq.~\eqref{eq:energy_rgg} to
\begin{equation}
\varepsilon_{ij}=
\ln{\frac{x_{ij}}{\kappa_i\kappa_j}},
\label{eq:energy_S1}
\end{equation}
and hence the connection probability in Eq.~\eqref{eq:conn_prob_FD} to
\begin{equation}\label{eq:conn_prob_S1}
p_{ij}=\frac{1}{1 + \chi_{ij}^{\beta}}=\frac{1}{1 + \left( \frac{x_{ij}}{\hat{\mu} \kappa_i \kappa_j} \right)^{\beta}}, 
\end{equation}
where $\kappa_i$ is the expected degree of node $i=1,\ldots,n$ in the ensemble~\cite{serrano2008similarity}, known as the $\mathbb{S}^1$ model (Fig.~\ref{part2_Fig_comm}a) and $\hat{\mu}$ a constant fixing the average degree. The values of parameters $\kappa_i$ can be either fixed or random, sampled from \emph{any} desired distribution. If they are sampled from the Pareto distribution~$\rho(\kappa)= (\gamma-1) \kappa_0^{\gamma-1}\kappa^{-\gamma}$, the resulting degree distribution is Pareto-mixed Poisson~\cite{hoorn2018sparse}, which for $k \gg 1$ is a power-law $P(k)\sim k^{-\gamma}$ observed in many real-world networks~\cite{voitalov2019scale}.
Clustering is still zero for $\beta\leq1$ and an increasing positive function of $\beta>1$.
The definition of edge energy in Eq.~\eqref{eq:energy_S1} combines the popularity (degrees $\kappa_i$) and similarity (distances $x_{ij}$) dimensions into a single measure, while the connection probability takes the gravity law form in Eq.~\eqref{eq:conn_prob_S1} decreasing with the similarity distance~$x_{ij}$ and increasing with the popularity product~$\kappa_i\kappa_j$.

The map
\begin{equation}\label{eq:kappa-to-y}
  \kappa\mapsto y=\kappa^2
\end{equation}
places all nodes~$i$ at coordinates~$(x_i,y_i)$, $x_i\in\mathbb{R}$, $y_i>\kappa_0^2$, in the upper half-plane model of the hyperbolic plane $\mathbb{H}^2$~\cite{cannon1997hyperbolic} that has a long history in relation to networks~\cite{gromov1987hyperbolic,matousek2002embedding,munzner1998exploring,aste2005complex,lohsoonthorn2003hyperbolic,jonckheere2007scaled,kleinberg2007geographic}. If $\rho(\kappa)$ is Pareto with $\gamma=3$, then nodes are distributed uniformly on the hyperbolic plane where the metric is $ds^2=(dx^2+dy^2)/y^2$. The group of distance-preserving isometries of the half-plane is isomorphic to the Lorentz group $SO(1,2)$. The Lorentz boosts, i.e., hyperbolic rotations in the $3$-dimensional Minkowski space, act on the upper half-plane as space-rescaling transformations $x \mapsto x'=\xi x$, $y \mapsto y' = \xi y$, where $\xi>0$~\cite{ratcliffe2006foundations}. The energies
\begin{equation}\label{eq:energy_S1_squaredmap}
  \varepsilon_{ij}=\ln\frac{x_{ij}}{\kappa_i\kappa_j}=\ln\frac{x_{ij}}{\sqrt{y_iy_j}}
\end{equation}
are thus manifestly invariant, and so is the model, with respect to rescaling Lorentz boosts, which form a noncompact subgroup of all isometries of the hyperbolic plane. However, the model is not invariant with respect to all isometries of the full Lorentz group since energy is not exactly a function of the hyperbolic distance, and so neither is the connection probability.

This problem is fixed in a slightly different but asymptotically equivalent model~\cite{krioukov2010hyperbolic} defined by the map 
\begin{equation}\label{eq:kappa-to-r}
\kappa\mapsto r=R-2\ln\kappa,
\end{equation}
where $R=2 \ln{(n/c)}$ and $c$ is the parameter controlling the average degree.  
This map places nodes~$i$ at polar coordinates~$(r_i,\theta_i)$, $\theta_i=2\pi x_i/n$, on the hyperbolic disc of radius~$R$ in the hyperboloid model of the hyperbolic plane with metric $ds^2=dr^2 + \sinh^2 r\,d\theta^2$~\cite{cannon1997hyperbolic} (Fig.~\ref{part2_Fig_comm}b). The edge energy becomes
\begin{equation}\label{eq:energy_H2}
  \varepsilon_{ij}=\frac{1}{2}\left(r_i+r_j+2\ln\frac{\theta_{ij}}{2}\right) \approx \frac{1}{2}d_{ij},
\end{equation}
where $\theta_{ij}$ and $d_{ij}$ are the angular and hyperbolic distances between the two nodes. The approximation holds for a fraction of node pairs that converges to~$1$ in the $n\to\infty$  limit~\cite{krioukov2010hyperbolic}. With this approximation the connection probability reads
\begin{equation}\label{eq:prob_H2}
  p_{ij} = \frac{1}{1 + e^{\frac{\beta}{2} (d_{ij} - R)}}.
\end{equation}

If $\rho(\kappa)$ is Pareto with $\gamma=3$ and $\beta\to\infty$, the model has the simplest formulation: sprinkle $n$ points uniformly at random over a hyperbolic disc of radius~$R$, and then connect all pairs of points located at distance $d_{ij}<R$ from each other. Fig.~\ref{part2_Fig_comm}b shows the equivalence between the $\mathbb{S}^1$ model and this $\mathbb{H}^2$ representation. Since energy is a function of the distance in~Eq.~\eqref{eq:energy_H2}, the model is fully Lorentz-invariant in the $n\to\infty$ limit for any~$\beta$.\\
\indent
The latent space in the model does certainly \emph{not} have to be the circle~$\mathbb{S}^1$. It can be \emph{any} compact homogeneous space of any curvature and dimension~$D$~\cite{boguna2020small}. The higher the dimension, the lower the clustering for the same value of~$\beta$~\cite{dall2002random,GarciaPerez2018}. Upon the hyperbolizing change of variables in Eqs.~(\ref{eq:kappa-to-y},\ref{eq:kappa-to-r}) these $D$-dimensional spaces turn into hyperbolic spaces of dimension $D+1$, while the edge energy becomes $\epsilon_{ij}=\ln{\frac{x_{ij}}{(\kappa_i\kappa_j)^{1/D}}}\approx d_{ij}/2$~\cite{boguna2020small}.\\
\indent 
The models have been also adapted to growing networks~\cite{papadopoulos2012popularity}, in which case the latent space is not hyperbolic but de Sitter space~$d\mathbb{S}^{1,D}$ with the same Lorentz group~$SO(1,D+1)$ of symmetries~\cite{krioukov2012network}, as well as to weighted networks~\cite{allard2017geometric}, multilayer networks~\cite{kleineberg2016hidden,Kleineberg2017}, and to networks with community structure~\cite{zuev2015emergence,muscoloni2018nonuniform,garcia-perez2018soft}. The model has yet to be extended to directed networks, as it is not clear how to reconcile the intrinsic symmetry of metric distances with asymmetric interactions among nodes.\\
\indent
The equivalence between the two models in Eqs.~(\ref{eq:energy_S1},\ref{eq:energy_H2}) is a reflection of the isomorphism between the Lorentz group~$SO(1,D+1)$ and the M\"obius group acting on sphere~$\mathbb{S}^D$ as the group of its conformal transformations. This isomorphism is a starting point of the  anti-de Sitter/conformal field theory (AdS/CFT) correspondence in string theory~\cite{maldacena1998large}.\\
\indent
It is important to reemphasize that the latent space in the described models can indeed be any compact homogeneous space of any dimension~$D$~\cite{boguna2020small}. This space can also be flat, or either positively or negatively curved. 
However, the effective space of the higher dimension~$D+1$ is always hyperbolic, i.e.\ negatively curved. What makes this space hyperbolic is the simple change of variables in Eqs.~(\ref{eq:kappa-to-y},\ref{eq:kappa-to-r}), mapping the expected degrees of all nodes to their $D+1$'th (radial) coordinates. Upon this change of variables, the probability of connections is always a function of hyperbolic $D+1$-dimensional distances only, Eqs.~(\ref{eq:energy_H2},\ref{eq:prob_H2}). However, the nodes are distributed uniformly according to the metric in this $D+1$-dimensional hyperbolic space, only if the degree distribution is a power law with exponent $\gamma=3$, i.e.\ only if $\rho(\kappa)$ is Pareto with $\gamma=3$. \\
\indent
Another very important observation is that, as was shown recently in~\cite{boguna2020small}, the described models are \emph{unique} latent-space network models that satisfy certain maximum-entropy requirements and that produce sparse heterogeneous uncorrelated small-worlds with non-zero clustering. At present, these models are also the only currently known class of network models that capture \emph{all} the following properties of many real-world networks: sparsity, self-similarity, small-worldness, heterogeneity, nonvanishing clustering, and community structure.\\
\indent
\textbf{\em Hyperbolic maps of real-world networks.} 
Given a real-world network, how to infer the coordinates of its nodes in its latent space? A collection of methods have been developed for this task.\\
\indent
Many generative-model-based methods perform statistical inference using Monte Carlo sampling and maximum likelihood estimation~\cite{boguna2010sustaining,papadopoulos2015network_mapping,papadopoulos2015network_geometry,blasius2018efficient_embedding}. Data-driven methods vary in flavor, ranging from unsupervised machine learning used, as in coalescent embedding~\cite{muscoloni2017machine} implementing nonlinear dimension reduction~\cite{alanis-lobato2016efficient}, to methods relying on the network community structure~\cite{wang2016hyperbolic,wang2016link}. Mechanistic-model-based methods rely on growing network automata to map the network while unfolding the similarity space~\cite{muscoloni2018minimum}.
If a generative model is a good description of a real network, and if Bayesian statistical inference based on such a model is done properly, such model-based methods tend to be accurate but slow, whereas data-driven methods tend to be faster but less accurate, so that hybrid methods have been also developed~\cite{alanis-lobato2016manifold,GarciaPerez2019}. They get initial coordinate estimates using machine learning techniques, and then refine the results via maximum likelihood estimation.
In general, the main challenge that hyperbolic network mapping methods face is the abundance of numerous local maxima in highly nonconvex likelihood landscapes. An efficient way to escape from these maxima thus boosting the inference accuracy is to shake up the system by adding to it decreasing levels of noise~\cite{kitsak2020link}, a method conceptually similar to simulated annealing.\\
\indent
The application of these methods to real-world networks allowed to investigate them at different resolutions using the latent-space renormalization group that unfolds the network in a self-similar multilayer shell of coexisting scales and their interactions~\cite{GarciaPerez2018}. The maps of real networks (Fig.~\ref{part2_Fig_comm}d--f) also revealed the existence of geometric communities~\cite{serrano2012uncovering,garcia-perez2016hidden}, and helped to decode mechanisms that govern network evolution, such as globalization, localization, and hierarchization driving the evolution of international trade~\cite{garcia-perez2016hidden}. Such maps have also shed light on many dynamical processes in real-world networks, showing, for instance, that cooperation in social networks is controlled more strongly by the latent-space organization than by highly-connected hubs in the system~\cite{kleineberg2017metric}.
Yet another class of applications of hyperbolic maps of real networks is link prediction. Since the connection probability in the described models is a decreasing function of the latent hyperbolic distance, the models predict that links are more likely to exist between hyperbolically closer pairs of nodes. Link prediction using hyperbolic geometry has been analyzed from different angles~\cite{muscoloni2018minimum,garcia-perez2020precision,kerrache2020scalable,kitsak2020link}. It appears to be particularly powerful when it comes to predicting links that are really difficult to predict~\cite{kitsak2020link}.
Finally, one of the most practical applications of mapping real-world networks to their latent geometries is the design of efficient routing protocols for the Internet~\cite{boguna2010sustaining} and for emerging Internet-of-Things telecommunication networks~\cite{voitalov2017geohyperbolic}. 
In the next few sections, we discuss some of the applications mentioned above.

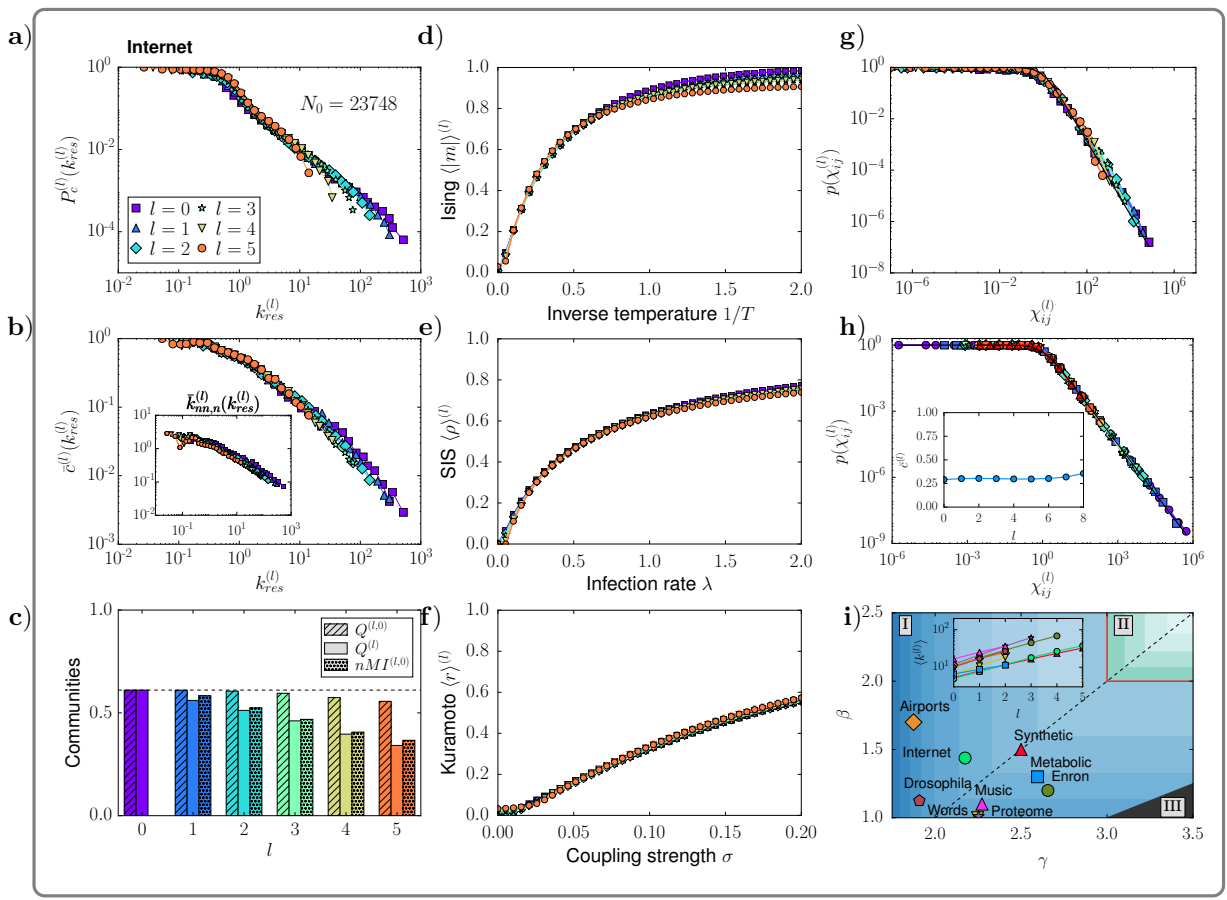
\begin{figure*}[t]
\tikzstyle{mybox} = [draw=gray, fill=white, very thick,
    rectangle, rounded corners, inner sep=5pt, inner ysep=10pt]
\tikzstyle{fancytitle} =[fill=cyan!25, text=black]
	\begin{tikzpicture}
    		\node [mybox] (box){
    		\includegraphics[width=0.9\linewidth]{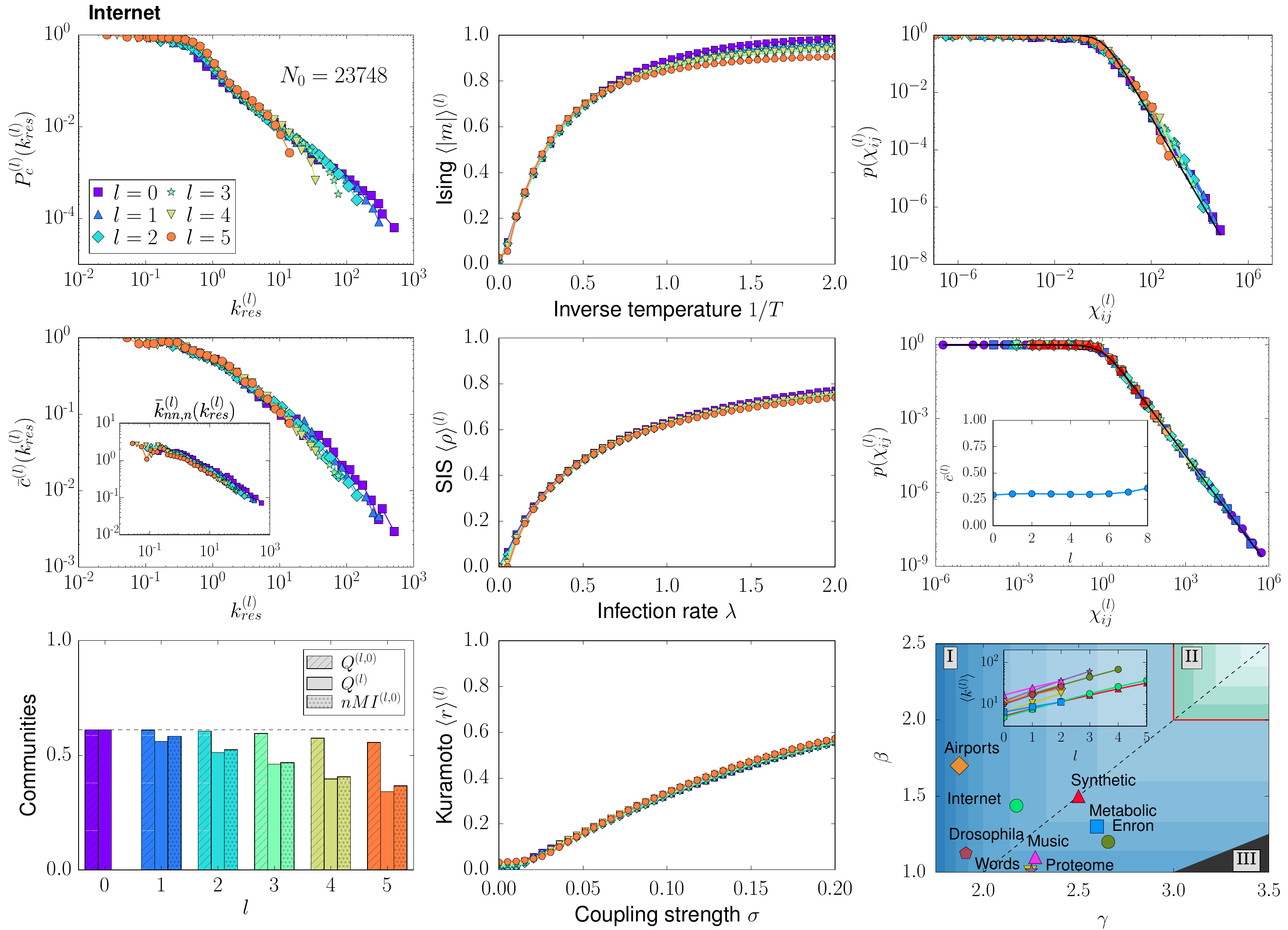}};
    		    \node[fancytitle,fill=none] at (-8.0,5.5) {\normalsize{\textbf{a})}};
	            \node[fancytitle,fill=none] at (-8.0,1.65) {\normalsize{\textbf{b})}};
	            \node[fancytitle,fill=none] at (-8.0,-2.2) {\normalsize{\textbf{c})}};
    		    \node[fancytitle,fill=none] at (-2.6,5.5) {\normalsize{\textbf{d})}};
	            \node[fancytitle,fill=none] at (-2.6,1.65) {\normalsize{\textbf{e})}};
	            \node[fancytitle,fill=none] at (-2.6,-2.2) {\normalsize{\textbf{f})}};
    		    \node[fancytitle,fill=none] at (2.95,5.5) {\normalsize{\textbf{g})}};
	            \node[fancytitle,fill=none] at (2.95,1.65) {\normalsize{\textbf{h})}};
	            \node[fancytitle,fill=none] at (2.95,-2.2) {\normalsize{\textbf{i})}};
\end{tikzpicture}
\caption{\footnotesize \textbf{Multiscale unfolding of networks' structure and function by geometric renormalization}. Self-similarity of the structural and dynamical properties of the multiscale 5-layered geometric renormalization shell of the Internet (a--g), and renormalizability of the $\mathbb{S}^1$ model (h--i) (courtesy of Ref.~\cite{GarciaPerez2018}). \textbf{a}) Degree distribution. \textbf{b}) Clustering spectrum. Inset: average nearest neighbors degree. \textbf{c}) Topological community structure. \textbf{d}--{\bf f}) Simulation of the Ising dynamics, SIS epidemic spreading dynamics and Kuramoto model for synchronization in the different layers of the shell. Results averaged over 100 simulations. d) Magnetization $\left<|m| \right>(l)$ of the Ising model as a function of the inverse temperature $1/T$. e) Prevalence $\left<\rho \right>(l)$ of the SIS model as a function of the infection rate $\lambda$. f) Coherence $\left<r \right>(l)$ of the Kuramoto model as a function of the coupling strength $\sigma$. \textbf{g}) Empirical connection probability in the Internet and the renormalized layers measured as the fraction of connected pairs of nodes as a function of ${\chi_{ij}}^{(l)} = R^{(l)} \Delta {{\theta}_{ij}}^{(l)}/(\hat{\mu}^{(l)} {\kappa_{i}}^{(l)} {\kappa_{j}}^{(l)})$. \textbf{h}) The same as in \textbf{g}), now for a synthetic $\mathbb{S}^1$ network with $N \approx 225000$ nodes, $\gamma = 2.5$ and $\beta = 1.5$. The black dashed line shows the theoretic curve given in Eq.~(\ref{eq:conn_prob_S1}). Inset: invariance of the mean local clustering along the flow. \textbf{i}) Real networks in the connectivity phase diagram of the $\mathbb{S}^1$ renormalization flow. The synthetic network above is also shown. Darker blue (green) in the shaded areas represent higher values of the exponent $c$ controlling the flow for the average degree $\langle k \rangle^{(l+1)} =  r^{c} \langle k \rangle ^{(l)}$. The dashed line separates the $\gamma$-dominated region from the $\beta$-dominated region. In phase I, $c > 0$ and the network flows towards a fully connected graph. In phase II, $c < 0$ and the network flows towards a one-dimensional ring. The red thick line $c = 0$ indicates the transition between the small-world and non-small-world phases. In region III, the degree distribution loses its scale-freeness along the flow. The inset shows the exponential increase of the average degree of the renormalized real networks $\left<k(l) \right>$ with respect to $l$.
}
\label{part2_Fig_renor}
\end{figure*}

\indent
\textbf{\em Geometric communities.}
The angular distribution of nodes in the described models is uniform. However, nodes in maps of real networks are clustered in regions defining geometric communities that were observed in many real networks, including the Internet~\cite{boguna2010sustaining}, metabolic networks in cells~\cite{serrano2012uncovering}, trade networks~\cite{garcia-perez2016hidden}, and brain connectomes~\cite{zheng2019a}. Non-overlapping communities can be detected in the geometric domain using purely geometric methods. One definition considers soft communities as groups of nodes in similarity space separated from the rest by angular gaps that exceed a certain critical value~\cite{serrano2012uncovering}. 
An alternative, known as the Critical Gap Method (CGM)~\cite{garcia-perez2016hidden}, finds the communities by changing the gap and selecting the soft community partition that maximizes the standard modularity measure~\cite{newman2006modularity}, Fig.~\ref{part2_Fig_comm}~c,f. These distance-based communities show strong correlation with groups defined by metadata, like geographical location of the Autonomous Systems in the case of the Internet~\cite{boguna2010sustaining}, biochemical pathways of reactions in metabolic networks~\cite{serrano2012uncovering}, or anatomical brain region in structural brain networks~\cite{Cacciola:2017ve,allard2018}. Geometric-based communities also show a significant overlap with topological-based communities, making the geometric nature of complex networks even more evident~\cite{garcia-perez2016hidden,faqeeh2018characterizing}.\\
\indent
Geometric communities affect degree-degree correlations and clustering spectrum in the network ensemble. These properties are not tunable but fixed by structural constraints in the vanilla version of the model described above with the homogeneous distribution of angular coordinates~\cite{boguna2004cut,catanzaro2004generation}. However, if these coordinate are not homogenously random but fixed to their heterogeneous values inferred in a real network with communities, the ensemble of random networks generated by the model using these heterogeneous inferred coordinates reproduces accurately the degree correlations and clustering in the real network~\cite{GarciaPerez2019,kitsak2020link}. This observation suggests that, to a great extent, degree correlations are a consequence of latent geometry coupled with inhomogeneous distributions of nodes in it.

\indent
\textbf{\em Navigability.}
One of the main strengths of the discussed latent-geometric network models lies in an explanation of the efficiency of this structure with respect to one of their most common functions, which is transport of information, energy, or other media, without the global knowledge of the network structure~\cite{boguna2009navigability}. 
Latent space guides navigation in the network based on distances between nodes in the space~\cite{boguna2009navigability}. That is, instead of finding shortest paths in the network---a computationally intensive combinatorial problem in a network that changes dynamically, such as the Internet~\cite{korman2006dynamic}---a transport process can be geometric, relying only on geodesic distances in the space (Fig.~\ref{part2_Fig_comm}h). Such processes are the more efficient and robust~\cite{muscoloni2019navigability}, so that the network is the more navigable, the smaller the~$\gamma$, and the larger the~$\beta$, defining a navigable parameter range to which many real-world networks belong~\cite{boguna2009navigability}. Networks in the hyperbolic model described above are nearly maximally efficient for such geometric navigation~\cite{krioukov2010hyperbolic}, which has recently been proven rigorously~\cite{bringmann2017greedy}.\\
\indent
The main reason behind this phenomenon is the existence, for any pair of nodes in hyperbolic networks, of shortest paths close to the corresponding geodesics in the underlying hyperbolic geometry (Fig.~\ref{part2_Fig_comm}h). 
Another critical factor is the existence of superhubs interconnecting all parts of the network, present as soon as~$\gamma<3$~\cite{krioukov2010hyperbolic}, in which case the networks are known to be ultrasmall worlds~\cite{cohen2003scale}. It was demonstrated in Ref.~\cite{boguna2009navigating} that navigation in hyperbolic networks with $\gamma<3$ can always find these ultrashort paths, and thus navigation in these networks is asymptotically optimal. The other way around, networks that are maximally navigable by design turned out to be similar to hyperbolic networks, and many real-world networks were found to contain large fractions of their maximum-navigability skeletons~\cite{gulyas2015navigable}, the human brain example shown in Fig.~\ref{part2_Fig_comm}j. Assuming that real-world networks evolve to have a structure efficient for their functions, these findings provide an evolutionary perspective 
on the emergence of latent geometry leading to structural commonalities observed in many different real-world networks. 

\indent \textbf{\em Renormalization and self-similarity.} Since networks in the discussed models are purely scale-invariant in the thermodynamic limit, they contain an infinite hierarchy of self-similar nested subgraphs induced by nodes with degrees exceeding a given threshold, as illustrated in Box B. 
This observation also applies to many real networks, where the average degree of the subgraphs increases as a function of the degree threshold~\cite{serrano2011percolation}. This property allows to prove easily the absence of percolation or epidemic thresholds in these networks~\cite{serrano2011percolation}, independently of the commonly used tree-like or scale-free assumptions~\cite{cohen2000resilience}. 
The proof is general and is solely based on a symmetry principle. 
Thus, it also applies to any phase transition whose critical point is a monotonic function of the average degree~\cite{serrano2011percolation}, like in SIS type epidemic spreading or in the Ising model, that in scale-free networks lack a healthy phase~\cite{pastor-satorras2001epidemic} or a disordered phase~\cite{bianconi2002mean,dorogovtsev:2002,leone2002ferromagnetic}, respectively.\\ 
\indent 
Self-similarity is also observed in the multiscale organization of networks, that can be explored at different resolutions by applying a geometric renormalization transformation~\cite{GarciaPerez2018} inspired by concepts from the real-space renormalization group in statistical physics~\cite{Wilson:1983,kadanoff,efrati2014real}. The method takes a different approach as compared to 
the shortest-path-distance RG discussed in Sec.~\ref{sec:fractal}, as it relies on distances in the similarity subspace to coarse-grain neighboring nodes into supernodes defining a new rescaled map, see Box C. The iteration of the transformation unfolds a network into a multiscale shell that progressively selects longer range connections revealing the coexisting scales and their interactions. Self-similarity under geometric renormalization is an ubiquitous symmetry in real-world networks, in good agreement with the prediction given by the renormalizability of the underlying $\mathbb{S}^1$ model~\cite{GarciaPerez2018}, Fig.~\ref{part2_Fig_renor}. This result suggests that the same connectivity law rules short and long range connections and operates at different length scales. From a practical point of view, applications include scaled-down network replicas and a multiscale navigation protocol that takes advantage of the increased navigation efficiency at higher scales. Interestingly, the structure of the human brain remains self-similar when the resolution length is progressively decreased by hierarchical coarse-graining of the anatomical regions, a symmetry that is well predicted by the geometric renormalization~\cite{zheng2019a}.

\begin{figure*}[!ht]
\centering
\includegraphics[width=\textwidth]{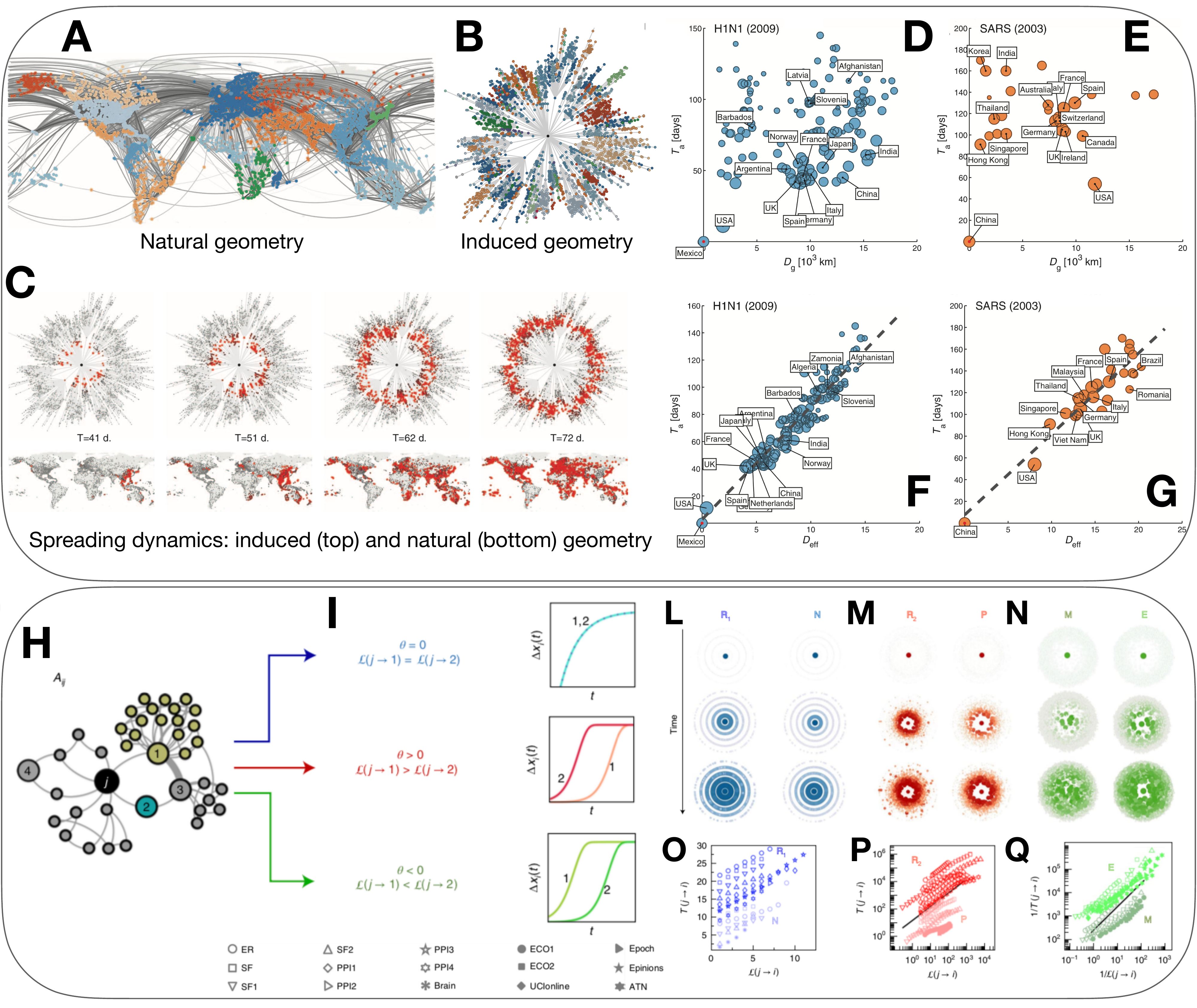}
\caption{\footnotesize\label{fig:effective_distance} (Top panel) \textbf{Geometry induced by spreading dynamics.} (A) Network representation of passenger flows along direct connections (edges) among airports (nodes) worldwide, with node color encoding geographic regions according to modularity maximization. The natural geometry of the network is given by its embedding into the physical space, with geodesic distance $D_{g}$. (B) The same nodes are embedded into a latent geometric space defined by effective distance $D_{\text{eff}}$ (see the text for details) accounting for the spreading dynamics on the top of the network. The center of this space is a node of the system, usually the one where the spreading originates from. (C) Evolution of a simulated disease spreading originating in Hong Kong (HKG): red symbols encode the prevalence. The top panels sketch the evolution in the latent, whereas the bottom panel shows the same dynamics in the natural space. When the spreading dynamics is depicted by exploiting the induced geometry, complex spatial patterns are mapped to homogeneous wave fronts propagating at constant effective speed. The relation between epidemic arrival times ($T_{a}$) and the two distances is shown for two real case studies, namely the spreading of H1N1 (D--F) and SARS (E--G). The relation is nonlinear when geographic distance is used (D--E) whereas it is nicely reproduced by a straight line when effective distance is considered. Reproduced with permission from Ref.~\cite{brockmann2013hidden}. (Bottom panel) \textbf{Geometry induced by universal temporal distance.} (H) A signal travels from vertex $j$ to the rest of the system exhibits different spreading patterns, captured by the universal temporal distance $\mathcal{L}(j\rightarrow i)$, (I) impacting different nodes (e.g., 1 and 2) in different ways depending on the type of dynamics (distance-limited, i.e. $\theta=0$; degree-limited, i.e. $\theta>0$; composite, i.e. $\theta<0$). (L--N) The homogeneous propagation of concentric wave fronts emerge from the analysis of a broad spectrum of synthetic and empirical systems. (O--Q) Propagation times $T(j\rightarrow i)$ of real-world signals are in agreement with the temporal distances $\mathcal{L}(j\rightarrow i)$. Reproduced with permission from Ref.~\cite{hens2019spatiotemporal}}
\end{figure*}

\section{Dynamic geometry of network processes}\label{sec:dynamics}

\begin{figure*}[!ht]
\centering
\includegraphics[width=\textwidth]{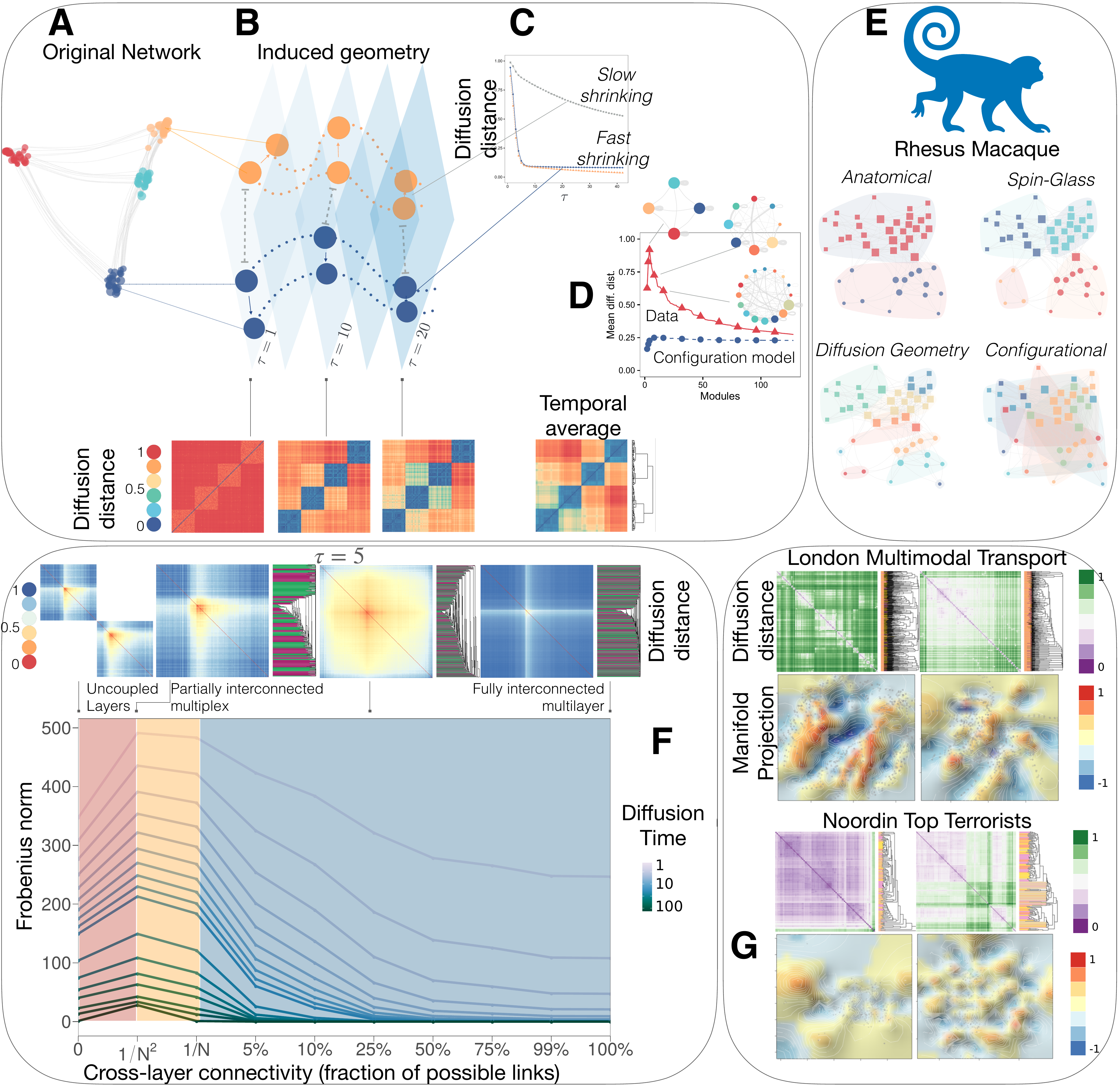}
\caption{\footnotesize\label{fig:diffusion_distance} (Top panels) \textbf{Diffusion geometry of complex networks.} (A) Euclidean embedding of a network with four clusters, corresponding to the latent diffusion geometry with Markov time $\tau=1$. (B--C) In the diffusion space, two nodes from the same functional cluster are (and keep) closer across time, $\tau$, than nodes belonging to different clusters. Diffusion-distance matrices corresponding to different times are shown in the bottom and allow to identify the underlying functional organization at different scales. (D) The functional modules maximizing the average diffusion distance define the mesoscale structure which favors the overall information exchange. 
The significance of this structure can be quantified by comparing against the result obtained from a configuration model preserving the degree distribution of the original data while destroying other correlations. (E) Diffusion geometry analysis of the anatomical connectivity (335 visual, 85 sensorimotor and 43 heteromodal) from 30 visual cortical areas and 15 sensorimotor areas in the Macaque monkey. Clusters identified by structural analysis of the connectome using the spin-glass approach are different, as the anatomical organization and the mesoscale organization obtained from the configuration model. \add{(Bottom panels) \textbf{Diffusion geometry of multilayer networks.} (F) Diffusion distance matrices (top) obtained from a two-layers synthetic network and a diffusive random walk dynamics. Colors in the dendrograms encode the original layer assignment of each node (only two colors, one for each layer): hierarchical clustering highlights the reorganization of nodes according to their function in the diffusion manifold. The main panel shows how the Frobenius norm of the supra-distance matrices changes for different regimes: from left to right, when two layers are not coupled together, when they are coupled by just one inter-layer link, when they form an interconnected multiplex connecting state nodes of the same physical node and when they tend to form a fully interconnected multilayer system. (G) As for (F), in the case of two empirical systems, the London public transportation network and the multiplex social relationships of Noordin Top terrorists. Multidimensional scaling applied to the distance matrices obtained from different random walk dynamics (here, multilayer maximum-entropy and physical walks with relaxation, see~\cite{bertagnolli2020multilayer} for details) allows to project the corresponding diffusion manifolds into $\mathbb{R}^3$. We show the surface on a place and encode the third dimension with colors to get a physical intuition about the functional organization of systems' units in the diffusion space.} Reproduced with permission from Ref.~\cite{dedomenico2017diffusion,bertagnolli2020multilayer}.}
\end{figure*}

Evidence for the existence of self-similarity and fractal properties characterizing the topology of empirical networks provided a solid ground for exploring the latent geometry of their structure in the hyperbolic space. The existence of a hidden geometry of network structure naturally led to question if a hidden geometry of network dynamics was plausible, aiming at identifying the latent space due to system function arising from the interplay between structure and dynamics. 
Note that network dynamics is broadly defined, including the dynamics \emph{of} vertex and edge creation or destruction---exemplified by, e.g., networks' growth processes or time-varying topologies~\cite{holme2012temporal}---as well as the dynamics of processes \emph{on} the network. 
\add{The hidden geometry induced by dynamics has been mostly explored for the latter, except for a few notable cases~\cite{muscolonilocal}}, with geometric tools 
finding fertile grounds also for unraveling hidden patterns lurking the complex behaviors of diverse network-driven processes. 
However, since it is possible to define multiple different dynamical processes on the network---e.g., epidemics spreading (Fig.~\ref{fig:effective_distance}) or random searches (Fig.~\ref{fig:diffusion_distance})---one expects not to find a unique latent geometry for the system's functioning: in principle, there could be as many hidden geometries as the number of plausible network-driven processes. 
In this universe of dynamical processes, the dynamics of information exchange has been \add{mostly modeled through diffusive processes, where the amount of matter being diffused is conserved, and spreading processes, in which the amount of matter being diffused is duplicated at each step (as in the spreading of viruses or ideas) and, consequently, it is not conserved. 
Such dynamics have been} successfully exploited to define novel metric or quasi-metric measures to \add{probe the corresponding latent geometry},
the difference being that, in the latter, the symmetry axiom is relaxed. \\
\indent 
Remarkably, \add{these classes of network geometry, induced by \emph{kinematic distances}, provide} results about a system's function that can not be obtained by geometric approaches discussed in the previous sections. 
An emblematic example concerns the mesoscale organization of interconnected components which exchange information during collective phenomena---say, coupled oscillators trying to synchronize or people with social relationships attempting to reach consensus---which can be characterized by mapping the interplay between structure and function to a geometric space induced by diffusion dynamics~\cite{dedomenico2017diffusion}. 
The resulting functional modules differ from the ones obtained by other geometric techniques such as functional modularity maximization~\cite{galvao2010modularity}, since the latter finds an optimal partition of the system while minimizing the possible number of modules of given topological size $\ell$, which in turns defines the characteristic distance between nodes within modules. 
Conversely, modules identified on the diffusion manifold at time $\tau$, determining the scale of dynamics, are characterized by groups of nodes which easily exchange information---for instance, in terms of random searches---within Markov time $\tau$, as we will see in the following.\\

\indent
\textbf{\em Geometry induced by resistance.} In first approximation, one can model information exchange in a network similarly to how currents flow in a circuit. This approach is probably among the first historical attempts to quantify a distance between nodes in terms of a (simple) dynamics. This \emph{resistance distance}~\cite{klein1993resistance} between two nodes $p$ and $q$ of a network is defined as $R_{pq}$ and it is calculated by assuming fixed resistors on each network edge: the corresponding circuit---under the further assumption that $p$ and $q$ are directly connected by a battery to allow electric currents to flow---allows for the calculation of effective resistances as
\begin{eqnarray}
R_{pq} &=& \Omega_{pp} + \Omega_{qq} - \Omega_{pq} - \Omega_{qp} ,
\end{eqnarray}
where $\boldsymbol{\Omega}=\mathbb{L}^{\dag}+N^{-1}\mathbb{U}$, being $N$ the size of the network, $\mathbb{U}$ a matrix with all entries equal to 1 and $\mathbb{L}^{\dag}$ the Moore--Penrose inverse of the Laplacian matrix of the network. On the one hand, this metric has been successfully used, for instance, to analyze specific isomers~\cite{babic2002resistance} and genetic differentiation~\cite{mcrae2006isolation}. On the other hand, it has been also shown that, 
for some classes of graphs, the resistance distance converges to the trivial thermodynamic limit $k_{p}^{-1} + k_{q}^{-1}$, where $k$ indicates the node degree~\cite{luxburg2010getting}, making this metric less useful for the analysis of most empirical systems. \\

\indent
\textbf{\em Geometry induced by communicability.} One of the first metrics based on dynamics is the communicability distance~\cite{estrada2012communicability,estrada2012complex}. 
This metric can be introduced by starting from the concept of 
{\em communicability} between two nodes $i$ and $j$ of a network, which is defined as $G_{ij}=\exp(\mathbf{A})_{ij}$, where $\mathbf{A}$ is the underlying adjacency matrix. 
From a mathematical point of view, communicability quantifies how well a pair of nodes exchanges information by all possible walks between them, giving more weights to the shortest ones, as it can be understood by
considering the Taylor expansion of the communicability matrix $\mathbf{G}$~\cite{estrada2012communicability,estrada2012complex}.

Communicability can be also well understood from a physical perspective. 
Consider a network where nodes are quantum harmonic oscillators and links are springs, and the system is submerged into a thermal bath with inverse temperature $\beta=1/kT$, being $k$ the Boltzmann constant. 
Then communicability provides a representation for the thermal Green's function of the system, indicating how a thermal oscillation propagates between nodes. 
The difference between the absorbed and transmitted excitation between two nodes due to such thermal disturbances is quantified by the \emph{communicability distance}~\cite{estrada2012complex}
\begin{eqnarray}
\xi_{pq}(\beta)&=&G_{pp}(\beta) + G_{qq}(\beta) - 2G_{pq}(\beta),
\end{eqnarray}
which allows to build an hyperspherical embedding of a complex network~\cite{estrada2014hyperspherical} at different temperatures. 
This graphical embedding allows to represent a geometry which is able to capture, for instance, spatial efficiency of networks~\cite{estrada2016communicability}, traffic flows in cities~\cite{akbarzadeh2018communicability} and constrained diffusion in coupled networks~\cite{estrada2018communicability} such as multilayer systems~\cite{de2016physics}. 
Since the topic in this respect is vast and beyond the scope of the present work, we refer the interested reader to Ref.~\cite{estrada2012physics} for a thorough review of the mathematical and physical properties of communicability distances in complex networks.
Let us also stress that, as the communicability distance, other important generalizations of the traditional shortest-path distance have been investigated in mathematics~\cite{chebotarev2011class,chebotarev2012walk}.\\

\indent
\textbf{\em Geometry induced by reaction-diffusion.} Another important class of geometry induced by network dynamics is obtained by considering a quasi-metric
as \emph{effective distance} can be used to gain insights about reaction-diffusion processes such as the spreading of infectious diseases through mobility networks~\cite{brockmann2013hidden,iannelli2016effective}. 
Given a network of geographic areas (e.g., airports) and edges encoding direct air traffic---in units of passengers per day---from node $i$ to  node $j$, let $F_{ji}$ indicate the corresponding mobility flow. 
Let also $P_{ij}=F_{ij}/\sum_{i}F_{ij}$ to quantify the fraction of this flow originating from node $j$ directed towards node $i$, defining the components of the connectivity matrix $\mathbf{P}$. 
Despite the structural complexity of the network, involving multiple and often redundant pathways for the transmission of contagion phenomena, the effective distance defined by
\begin{eqnarray}
\delta_{ij}=1-\log P_{ij},
\end{eqnarray}
reveals hidden geometric patterns where the dynamics of epidemics spreading is elegantly mapped into the propagation of wavefronts with an effective speed. 
This latent dynamic geometry can be used to better predict the arrival times of empirical contagion processes in distinct geographic areas and to reconstruct, with reasonable accuracy, the origin of outbreaks (Fig.~\ref{fig:effective_distance}). 
\\
\indent 
Similarly, the 
collective dynamics of different signal spatio-temporal propagation on networks can also be unraveled by looking at the geometry induced by their dynamics upon the definition of the 
\emph{universal temporal distance}~\cite{hens2019spatiotemporal}
\begin{eqnarray}\label{eq:barzel}
\mathcal{L}(j\rightarrow i)=\min\limits_{\Pi(j\rightarrow i)}\left\{\sum\limits_{p\in\Pi(j\rightarrow i), p\neq j} S_{p}^{\theta}\right\}.
\end{eqnarray}
\noindent 
In Eq.~\ref{eq:barzel}, $\Pi(j\rightarrow i)=j\rightarrow q\rightarrow ... \rightarrow i$ indicates the shortest-path from the origin $j$ to the destination $i$, while the delay $\tau_{p}$ in signal propagation occurring on each node $p\in\Pi(j\to i)$ is assumed to scale as $\tau_{p}\sim S_{p}^\theta$, where $S_{i}=\sum_{k}A_{ik}$ a node weighted degree 
and $\theta=-2-\Gamma(0)$ a parameter determined by the system's dynamics.
As shown in Ref.~\cite{hens2019spatiotemporal}, this metric yields excellent predictions about the actual propagation times $T(j\rightarrow i)$ for a range of nonlinear dynamic models (Fig.~\ref{fig:effective_distance}h,i), further showing that, despite their diversity, disparate propagation patterns actually condense into three highly distinctive dynamic regimes (highlighted in blue, red and green in Fig.~\ref{fig:effective_distance}l--q) characterized by the interplay between network paths, degree distribution and the interaction dynamics.\\
\indent
In the same spirit, spreading processes on noisy geometric networks~\cite{barthelemy2011spatial,bonamassa2019critical} 
have been recently investigated to understand how contagion dynamics is driven by the underlying topology~\cite{taylor2015topological}.
Noisy geometric networks provide a suitable framework to model systems characterized by both 
short (i.e.,\ large-world) and 
long-range (i.e.,\ small-world) interactions among nodes. Contagion maps, whose manifold structure reflects the interplay between local and non-local structure with the epidemic spreading process, provide a suitable tool to recover the geometric features of a network's underlying manifold and describe wavefront propagation in the corresponding geometric space. This geometric framework allows to gain physical insights on contagions
by exploiting computational topology and to identify low-dimensional structure in complex networks~\cite{taylor2015topological}.\\

\indent
\textbf{\em Geometry induced by random search.} More recently, random walk dynamics has been proposed to define a \emph{diffusion distance} between pair of nodes. 
Let $\boldsymbol{\mathcal{L}}=\mathbf{I}-\mathbf{T}$ indicate the normalized Laplacian matrix, with $I_{ij}$ the Kronecker delta and $T_{ij}$ the probability for the random walker to move from node $i$ to node $j$. Let $\mathbf{e}_{i}\equiv(0,0,...,1,...,0)$ be the $i$--th canonical vector in the Euclidean space with dimension $N$, the size of the system. The evolution over time $\tau$ of the probability to find the walker in any node is described by a master equation~\cite{masuda2017random}, whose solution is given by $\mathbf{p}(\tau | i)=\mathbf{e}_{i}\exp{(-\tau \boldsymbol{\mathcal{L}})}$ when the initial condition is $\mathbf{p}(0)=\mathbf{e}_{i}$, i.e., the walk's origin is in node $i$ with probability 1. The hidden geometric space induced by the random walker's Markov dynamics~\cite{dedomenico2017diffusion} is characterized by the diffusion distance between nodes $i$ and $j$, defined by
\begin{eqnarray}
d_{i,j}(\tau)=\norm{\mathbf{p}(\tau|i) - \mathbf{p}(\tau|j)},
\end{eqnarray}
providing, among others, the starting point to build diffusion maps which are widely adopted for low-dimensional embedding of high-dimensional data~\cite{coifman2005geometric}. 
Two nodes are close in their latent diffusion space if connected by multiple pathways which facilitate information exchange in less than $\tau$ steps. 
As a direct consequence, the mesoscale functional organization of the network is mapped into spatial clusters in the corresponding diffusion manifold, with Markov time playing the role of a multi-resolution parameter, i.e. a temporal length scale playing the dynamic analogue of the shortest-path-distance in Sec.~\ref{sec:fractal} or the similarity metrics of Sec.~\ref{sec:hyperbolic}.
Increasing values of $\tau$, in fact, allow to identify functional hierarchies 
at multiple temporal resolutions, 
whose persistence 
across time identifies the mesoscale structure which favors the overall information exchange, providing the best coarse-groaning of the system in functional modules (Fig.~\ref{fig:diffusion_distance}(A--E)). \add{Micro-, meso- and large-scale structures can be probed for small, increasing and large $\tau$, respectively.}
Geometry induced by diffusive processes allows to gain physical insights about collective phenomena in structured populations, by establishing a formal relationship with complex dynamics responsible for synchronization in the metastable state and emergence of consensus. \\
\indent
The recent application to anatomical connectivity within and between visual cortical and sensorimotor areas in Macaque brain reveals a hierarchical functional organization of cortical units, not identified by existing methods and not compatible with null models~\cite{dedomenico2017diffusion}. 
The network embedding in a geometric space induced by diffusion distance, together with statistical data depth, allows for the statistical and most natural generalization of the concept of median to the realm of complex networks, with the advantages for defining the centre of the system and percentiles around that centre to identify vertices which are socially or biologically relevant~\cite{Bertagnolli2019}. 

\add{More recently, the framework has been extended to the realm of multilayer networks~\cite{bertagnolli2020multilayer}, capitalizing on the generalization of different random walk dynamics to multilayer systems~\cite{dedomenico2013mathematical,dedomenico2014navigability,de2016physics}. In fact, it has been show that layer-layer topological correlations might alter information exchange among state nodes, while the presence of different inter-layer connectivity patterns might lead to distinct geometric regimes: i) when the fraction of inter-links is small, flow is segregated within layers and the diffusion manifold consists of two well separated sub-manifolds, corresponding to the functional organization of each layer separately, connected by weak geometric pathways; ii) when the fraction is sufficiently high the flow is integrated, new geometric pathways are made available to information to be exchanged across layers and those sub-manifolds mix up. Different multilayer diffusion manifolds have been used to better understand the functional organization of multimodal transportation and multiplex social systems (Fig.~\ref{fig:diffusion_distance}(F--G)).} 

\section{Discussion and Outlook}\label{sec:discussion}

Network geometry provides tools and methods complementary to those inspired by classical statistical mechanics in network science. 
Despite its relatively recent inception, network geometry has already served as a remarkably successful pathway to harness the observable and hidden forms of symmetries in many real-world systems, leading to a wealth of discoveries of both theoretical and practical importance. 
The selection of milestones presented in Sections~\ref{sec:fractal}--\ref{sec:dynamics} offers a mature viewpoint from where to ponder on some emerging research directions and challenges 
ahead.\\
\indent
{\bf \em Fractal geometry of network structure} has enabled the study of many relevant aspects related to the self-similarity observed in various complex networks, casting their understanding under the three pillars~\cite{stanley1999scaling} of scaling, universality and renormalization. 
If on the one hand the simplicity of this approach has helped disclosing fundamental insights about notoriously hard problems on complex media---such as transport~\cite{gallos2008scaling,condamin2007first}, functional modularity~{\cite{gallos2012conundrum,galvao2010modularity} or evolution~\cite{song2006origins,jin2013evolutionary}}---on the other one it has raised profound questions to ponder on. \\
\indent 
A pragmatic example involves the very same notion of self-similarity and fractality in complex networks. 
Although the traditional fractal theory does not distinguish between fractality and self-similarity, these two properties can be considered to be distinct under the lens of the shortest-path distance RG~\cite{gallos2007review}. 
In fact, while both fractal {\em and} pure small-world structures are statistically self-similar under the shortest-path distance RG (see, e.g.~Fig.~S6 in Ref.~\cite{song2006origins} and discussions therein), only the former family features a well defined set of (finite) fractal exponents. 
The divergence of the box-counting dimension $d_B$ in small-world networks (Fig.~\ref{fig:1}c) can be then interpreted (using a field-theory jargon) as an ``ultraviolet'' limit above which the fractal geometric approach based on the shortest-path RG fails in quantifying their self-similar symmetry.
In this perspective, identifying suitable embeddings or duality transformations~\cite{bloch1946methods} capable of regularizing this ``small-world divergence'' is a theoretical open challenge that could raise the opportunity of designing a unified geometric framework for the study of fundamental problems like transport~\cite{villani2008optimal}, evolution~\cite{zheng2019b}, navigability~\cite{boguna2009navigability} or RG-based classification of networks' universality classes~\cite{rozenfeld2010small}. 
In this respect, latent-geometric approaches or the adoption of generalized topological metrics and embeddings~\cite{estrada2012communicability} could offer hints for solutions. 
Hyperbolic geometry, in particular, shares profound connections with self-similar metric spaces~\cite{gromov1987hyperbolic,nekrashevych2003hyperbolic,nekrashevych2005similar}, suggesting the possibility that the fractal exponents featured by self-similar networks may have suitable counterparts in their corresponding latent spaces and could further be extended to pure small-world structures.\\ 
\indent 
On a more fundamental level, understanding how the dynamical properties of network growth processes influence their asymptotic self-similar patterns is an intriguing and yet unexplored territory in the study of network geometry. 
The process of zooming-out induced by the network RG transform finds its (statistically equivalent) inverse in dynamics~\cite{song2006origins,zheng2019b}, so that the varying structures observed at increasing length scales correspond to the evolution of certain dynamical variables (Fig.~\ref{fig:self2}\textbf{a}). 
A profound bridge between self-similarity and ergodicity has been well explored in mathematics~\cite{furstenberg2014ergodic}, showing that the notions of fractal dimensions and self-similarity can be interpreted in terms of ergodic averages of some appropriate measure preserving dynamical systems~\cite{furstenberg2003markov}. 
In the simple case of growing trees, it has further been proved~\cite{furstenberg2014ergodic} that this connection is a consequence of the explicit dependence of the fractal dimension on the growth rates ruling the system's evolution.
A perfectly analogue result was reported in Ref.~\cite{song2006origins}, where the fractal exponents 
$d_B,\,d_k,\,d_M,\,\dots$ characterizing the self-similar structure of the SHM model in the thermodynamic limit were proven to depend only on the process' growth rates (Fig.~\ref{fig:self2}\textbf{b}). 
A relevant and highly challenging question, in this respect, is to understand weather the SHM model and/or other more popular growth processes~\cite{dorogovtsev2002evolution,song2006origins,zheng2019b}, can be themselves interpreted as ergodic dynamical systems with respect to some suitable invariant measure. 
Besides its theoretical relevance, finding viable directions to tackle this overarching problem could help establishing fundamental connections among the static and the dynamic facets of network geometry.\\ 
\indent
{\bf \em Hyperbolic geometry of network latent spaces} impacted 
areas as diverse as mathematics, neuroscience, and machine learning. Random graphs have a long research history in graph theory, probability, and adjacent areas in applied mathematics and theoretical computer science (TCS). 
Given that hyperbolic networks turned out to be the first popular ensemble of random graphs reproducing not only inhomogeneous degree distributions but also nonvanishing clustering, small-worldness, self-similarity, and modularity observed in many real-world networks, this ensemble attracted significant research attention in mathematics and TCS, where many basic and advanced properties of random hyperbolic graphs have been (re)derived rigorously, see for instance~\cite{gugelmann2012random,fountoulakis2015geometrization,candellero2016clustering,bode2016probability,bringmann2017greedy,abdullah2017typical,blasius2018cliques,fountoulakis2018law,friedrich2018diameter,kiwi2018spectral,bringmann2019geometric,muller2019diameter}.\\
\indent 
In neuroscience, geometric navigation discussed in Sec.~\ref{sec:hyperbolic} offers a possible explanation and a mechanism for the routing of information in the brain.
This hypothesis has been investigated at different depths and from different angles~\cite{heuvel2012high,goni2013exploring,fornito2013graph,misic2014communication,roberts2016contribution,avena-koenigsberger2017path,seguin2018navigation,tadic2018origin,avena-koenigsberger2019spectrum,wang2019synchronization}. 
Yet, geometric navigation is effective only when the network topology is congruent with the underlying latent geometry so that following geodesic paths in the latent space is equivalent to navigating through topological shortest paths. 
This seems to be the case for structural brain networks whose many structural and navigability properties were shown to be
well described by the $\mathbb{S}^1$/$\mathbb{H}^2$ geometric network model~\cite{Cacciola:2017ve,allard2018}, where the same connectivity laws apply both to short- and long-range connections and at different scales~\cite{zheng2019a}. 
This suggests that simplicity might be one of the main organizing principles of human structural brain networks at least at the macroscale level that has been found to display a self-similar architecture across different anatomical length scales in good agreement with the discussed geometric models~\cite{zheng2019a}, as opposed to traditional approaches that describe brain connectivity using Euclidean geometry~\cite{betzel2015generative,horvat2016spatial,stiso2018spatial}.
Euclidean distances do certainly play a role in the brain. However, they are not the only factor determining similarity, and thus connectivity, between brain regions~\cite{allard2018,zheng2019a}. More recently, data-driven dimensional reduction techniques~\cite{allen2015intrinsic,Cacciola:2017ve} and local curvature measures~\cite{farooq2019network} have been proposed as alternative geometric descriptions that avoid the definition of connectivity laws. It still remains an open question how the hyperbolic geometry of the brain relates to the optimization principles that Ram\'on y Cajal hypothesized about 100 years ago as the two forces ruling the evolution of mammalian brain connectivity: minimizing wiring costs and maximizing conductivity speed~\cite{assaf2020conservation}.\\
\indent
At the same time, primarily after Ref.~\cite{nickel2017poincare}, hyperbolic spaces have ignited vigorous research activity in machine learning in many different settings and tasks including embedding graphs and other data, such as images and texts, as well as in the design of neural networks, attention networks, knowledge graphs, and matters alike, with applications ranging from data classification, image recognition, and natural language processing, to link prediction and scalable recommender systems~\cite{dhingra2018embedding,ganea2018hyperbolic_entailment,ganea2018hyperbolic_neural,nickel2018learning,ovinnikov2018poincare,sala2018representation,gulcehre2019hyperbolic,chami2019hyperbolic,liu2019hyperbolic,suzuki2019hyperbolic,tifrea2019poincare}. Overall, the main flavor of these results confirms one of the main points in Ref.~\cite{krioukov2010hyperbolic}: compared to Euclidean geometry, hyperbolic geometry appears to be a better (embedding space) model for highly heterogeneous networks and other data.\\
\indent
In terms of open questions, perhaps one of the most common ones is how to tell whether a given (real-world) network has a latent (hyperbolic) space. This question is a variant of the more general question of how to tell whether a given model is a good model for a given network. Such questions can never find positive answers since one can never know for sure whether any given network has typical values of all possible network properties in any given model, simply because the number of such properties is infinite~\cite{orsini2015quantifying}. One can always check if any finite collection of properties of the network have the typical values in the model, and as soon as an atypical property is found, one may raise doubts how good the model is for the network at hand. Such a finding does not always render the model useless. For instance, clustering is zero in stochastic block models, and it is nonzero in real-world networks, but this fundamental mismatch does not appear to diminish much the interest to stochastic block models, thanks to their simplicity and tractability.
\indent
As mentioned in Sec.~\ref{sec:hyperbolic}, the latent-space models described there reproduce most of the important
properties of real-world networks. 
What this means is that any network that has these properties can be mapped to a latent space of any dimension,
yielding some nontrivial results. What space dimension one should choose for the embedding~\cite{muscoloni2017machine,gu2020defining}, is an interesting open problem. It calls to identify a network property that would depend on the space dimension in a known way. At present, it is known that clustering is a decreasing function of dimension~\cite{dall2002random,GarciaPerez2018}, but it is also a decreasing function of temperature~\cite{serrano2008similarity,krioukov2010hyperbolic}, so that the value of clustering by itself cannot tell us the likely value of the space dimension. In the lack of such understanding, and given that space in hyperbolic geometry expands exponentially fast in any dimension so that the ``crowding problem'' is never a problem there, there are no reasons not to choose the simplest case~$D=1$ for the embedding, unless overparametrization may be beneficial, as in deep learning~\cite{chaudhari2019entropy}.\\
\indent
Another important open problem is to identify a collection of network properties that are not only necessary but also sufficient conditions for latent geometricity. This problem is not about any given network, in which case it can never be solved for the reasons above, but about network models. For instance, the question about whether clustering is such a property can be formulated as follows: does a model that reproduces clustering but produces otherwise maximally random networks is equivalent to a latent-space model? Such questions cannot be answered without additional stringent assumptions and requirements, such as the requirement of maximum entropy. The first steps in this directions have been made in~\cite{krioukov2016clustering,boguna2020small}.\\
\indent
Other open problems, which are very interesting from the practical perspective, are to study the coupling of dynamical processes with the latent space and to generalize latent space network models to temporal networks~\cite{holme2012temporal}. It has been observed that clusters in the underlying metric space emerge in evolutionary games on scale-free networks~\cite{kleineberg2017metric}. Other processes could present geometric patterns in their dynamical states as well. A key point here is that essentially all real-world networks are highly dynamic at different timescales, suggesting that the positions of nodes in the latent space cannot be really fixed but must constantly change. 
The development of dynamic latent space models is, however, very challenging, because
there is no consensus or even general understanding concerning the exact set of requirements to such models. The main high-level problem here is that there are too many ``degrees of freedom'' in designing these models, that is, arbitrary choices can be made, and there are no generally agreed guiding principles concerning what choice is better or worse.\\
\indent
At a more fundamental level, the fact that hyperbolic networks are Lorentz-invariant in the thermodynamic limit suggests to re-examine the role of probabilistic symmetries~\cite{kallenberg2005probabilistic} in the theory of graph limits~\cite{lovasz2012large}. Traditionally the main symmetry of interest there has been exchangeability~\cite{orbanz2015bayesian}. This is the requirement that the probability of a graph in an ensemble does not depend on how nodes in the graph are labeled, reminiscent of gauge invariance in physics. This requirement is really stringent for deep statistical reasons~\cite{orbanz2015bayesian}, but it is also easy to appreciate intuitively: if node labels $1, \ldots, n$ are just random ``coordinates'' used to represent an otherwise unlabeled graph as an adjacency matrix, then the probability of the graph in the ensemble cannot depend on these meaningless coordinates. The problem is the Aldous-Hoover theorem~\cite{aldous1981representations,hoover1979relations}, stating that the thermodynamic limit of any exchangeable sparse graphs is exactly empty. This is why different notions of exchangeability-like probabilistic symmetries have been recently investigated for sparse graph limits~\cite{caron2017sparse,veitch2015class,borgs2019lp,janson2017edge}. Because of the Aldous-Hoover theorem, they all depart from the $1, \ldots, n$ ``coordinate system'' for node labels, and rely on different systems of node or edge labels. The Lorentz invariance of hyperbolic networks suggests that the labels of nodes can be indeed their coordinates in a latent space, with exchangeability replaced by invariance with respect to the space isometries. The hope is that such ensembles may have some interesting and tractable thermodynamic limits.\\

\indent
{\bf \em Dynamic geometry of network processes.} 
 While the results on observable and latent geometries of networks are well established, the geometry of network-driven processes is in its infancy, with promising theoretical developments and important applications to different fields. 
From the global spread of rumors and opinions in socio-technological systems to the global spread of innovations and epidemics, combining dynamics with the self-similar structure of complex networks and their latent geometry results in heterogeneous processes which cannot be easily understood when investigated in the Euclidean space where they are often embedded.  However, when the same dynamical processes are analyzed through the lens of the geometries they induce, one often discovers simple and elegant arguments to better understand the complex spatio-temporal patterns observed in a broad spectrum of complex systems.
These results make the geometry induced by network-driven processes perhaps the most suitable framework for several practical applications, ranging from predicting the time course of dynamics for forecasting and control of spreading processes to locating their origin. Indeed, diffusion geometry defines a class of models that have the desirable advantage of being mathematically tractable and can be easily interpreted.\\
\indent
The research program for the future is broad, with many open challenges of theoretical and practical relevance. \add{On the one hand}, as the topological organization of empirical complex systems can be characterized in terms of hierarchies~\cite{ravasz2003hierarchical,corominas2013origins} and mesoscale structures such as bow-tie~\cite{newman2001random,dorogovtsev2001giant}, $k$--cores~\cite{seidman1983network,dorogovtsev2006k} and core-periphery~\cite{holme2005core,rombach2014core}, it will be interesting to identify and characterize their functional counterparts in terms of diffusion geometry. Here, the main problem is to define mesoscale objects such as functional giant components and functional cores. 
\add{On the other hand, further research is needed to better understand the deep relationships among the different aspects of network geometry reviewed here. A promising step towards the direction of building a theoretical bridge is provided by the concept of communicability, which can be understood in purely combinatorial terms as an effective pathway -- weighting in a very specific way the contributions of walks of different lengths -- between nodes in the network space. In this regard, it is intimately related, for instance, to the concept of geometrical navigability, which is determined by selecting topological paths that follow geodesics in the latent space. In general, those paths happen to be topological shortest-paths when networks are sufficiently congruent with the latent space.}
Notice that congruency can be defined in different ways, not only in terms of greedy shortest paths, as done in~\cite{boguna2010sustaining,krioukov2010hyperbolic,papadopoulos2010greedy,allard2018}, but also taking into account all topological shortest paths~\cite{cannistraci2020geometrical}.\\ 
\indent
As for further open challenges, developing RG techniques in the space induced by diffusion distances is a fundamental open problem, whose solution could shed light on the self-similar symmetries of dynamical processes evolving on the top of complex networks. This could further pave the way to the analysis of coexisting temporal scales due to the interplay between structure and dynamics.
Finally, a natural development of geometries induced by network-driven processes is the identification and characterization of a more general framework where diffusive dynamics are replaced by more complex ones. This advance would improve our understanding of complex dynamical processes and has the potential to enhance the control and forecasting of the evolution of empirical systems.

\indent
{\bf \em Other flavors of network geometry.} The spectacle of flavors of network geometry is by no means limited to the topics addressed in details above. 
Other significant direction of research in network geometry have emerged through the years. Recent advances and open challenges in some of those areas are briefly reviewed below.\\
\indent
{\it Geometrogenesis.} Perhaps one of the most fundamental open problems in network geometry is that of geometrogenesis, which is the emergence of continuous geometric spaces from discrete combinatorial rules. 
The study of emergent geometries is recently gaining large momentum~\cite{wu2015emergent,mulder2018network} due to its intimate connections with longstanding combinatoric problems in several approaches to quantum gravity, like {\em causal sets}~\cite{rideout2009spacelike,dowker2011spacetime}, {\em quantum graphity}~\cite{Konopka2008} and {\em causal dynamic triangulations}~\cite{ambjorn2005}.
\indent
In loop quantum gravity (LQG), for example, the basis of states is formed by spin networks which have support on a graph, determining a sort of quantum geometry where the \emph{intrinsic} geometry---consisting of quanta of space---is discrete and the \emph{extrinsic} curvature is fuzzy because of the Heisenberg's uncertainty principle~\cite{rovelli2010geometry}. The main challenge is to assign a classical geometrical interpretation to such states. Recent advances in this direction are based on operators which quantize both scalar and mean curvature when spin network edges run within the surfaces of the quantized geometry~\cite{gruber2018geometry}. In quantum graphity~\cite{Konopka2008}, the space is a dynamical graph evolving under the action of a Hamiltonian. In causal dynamical triangulation (CDT)~\cite{Ambjorn2011}, a non-perturbative path integral approach is used to build a connection with Hovrava-Lifshitz gravity in $2+1$ dimensions~\cite{Sotiriou2011}. Spectral dimension, defined as the scaling exponent of the average return probability of diffusion processes (e.g., random walks), is used in CDT to measure the effective dimension of the underlying geometry~\cite{loll2019quantum} and can provide an interesting bridge to geometry induced by network-driven processes, where one expects that this geometry characterizes the underlying diffusion manifold. More recently, a model~\cite{kelly2019assembly} where random graphs dynamically self-assemble into discrete manifold structures has been proposed as an alternative to approaches based on simplicial complexes and Regge calculus. The Ollivier curvature, defined for generic graphs---and similar in spirit to Ricci curvature---is used to discretize the Euclidean Einstein-Hilbert action and to provide a new ground for emergent time mechanisms~\cite{kelly2019assembly}.\\
\indent
In network science, a step forward in addressing the geometrogenesis problem has been made in Refs.~\cite{wu2015emergent,bianconi2015complex_quantum_network_manifolds,bianconi2016network,bianconi2017emergent,mulder2018network} by defining models of growing random graphs and simplicial complexes. In a wide range of parameters, these models lead to an effective preferential attachment and, thus, heterogeneous degree distributions~\cite{bianconi2015complex_quantum_network_manifolds}. Besides, some growth processes can be mapped to trees, leading to emergent hyperbolic geometry in the resulting graphs~\cite{bianconi2017emergent}. 
For a more exhaustive introduction to the topic we refer the interested reader to the recent review by Mulder and Bianconi~\cite{mulder2018network}.\\
\indent
{\it Graph curvature.} Curvature is one of the most basic geometric notions, a key player in the Einstein-Hilbert action, whose least-action variation leads to Einstein's equations in general relatively. It is thus not surprising that graph curvature appears in many flavors of geometrogenesis and combinatorial approaches to quantum gravity~\cite{kelly2019assembly}. More surprising is that there is not one but many successful attempts to port the notion of curvature to the realm of networks, resulting in many nonequivalent definitions of graph curvature~\cite{higuchi2001combinatorial,aste2005complex,ollivier2009ricci,lin2011ricci,sreejith2016forman,sreejith2017systematic,samal2018comparative,prokhorenkova2020global}. Unfortunately, none of these definitions of graph curvature is rigorously known to converge to any traditional curvature of smooth space in the continuum limit of any random graph ensemble, with the only exception of Ollivier curvature of random geometric graphs recently shown to converge to Ricci curvature in any Riemannian manifold~\cite{hoorn2020ollivier}.\\
\indent
Nevertheless, many of these graph curvature definitions have recently found applications in diverse applications ranging from differentiating cancer networks~\cite{sandhu2015graph} to the characterization of human brain structural connectivity~\cite{farooq2019network}, to mesoscale characterization of complex networks based on Ollivier-Ricci curvature~\cite{sia2019ollivier} and Ricci flow~\cite{ni2019community}. The topic is of much interest and receiving increasing attention, since combinatorial-based notions of network curvature have disclosed profound connections between network measures such as the graph Laplacian and exquisitely geometrical quantities such as the Laplace-Beltrami operator in Riemannian manifolds~\cite{robles2007riemannian,majid2013noncommutative} or the Fisher-Rao metric in information geometry~\cite{franzosi2016riemannian}.\\
\indent
An important definition of global graph curvature that applies to networks without a change, is Gromov's $\delta$-hyperbolicity~\cite{gromov1987hyperbolic}, which has been measured for a variety of real-world networks~\cite{narayan2011large,albert2014topological,borassi2015hyperbolicity,tadic2019functional}. It is a rough measure of how far a metric space is from a tree~\cite{chatterjee2019average}. It applies to any metric space, including the metric spaces of shortest path distances in Section~\ref{sec:fractal} and the latent spaces in Section~\ref{sec:hyperbolic}. Any hyperbolic space is also $\delta$-hyperbolic~\cite{buyalo2007elements}, but a network is called $\delta$-hyperbolic if its shortest path metric space is such. This terminology often causes bad confusion because networks in the hyperbolic latent-space models in Section~\ref{sec:hyperbolic}---that are often called \emph{random hyperbolic graphs}---are actually unlikely to be $\delta$-hyperbolic because their two different limits are not $\delta$-hyperbolic~\cite{shang2012lack,shang2013hyperbolicity}. However, at present it is not exactly known how $\delta$-hyperbolic the random hyperbolic graphs are, another open problem.\\
\indent
{\it Topological data analysis.} \add{More recently, another interesting bridge with topology has been established by generalizing networks to higher dimensions via simplicial complexes, allowing} for the application of persistent homology methods~\cite{aktas2019persistence} from topological data analysis (TDA)~\cite{patania2017topological}. Persistent homology relies on the filtration of a simplicial complex to uncover topological features that recur over multiple scales and are thus likely to represent some true features of the underlying space. The TDA in general and persistent homology in particular are vigorous research areas in data science that found applications in a variety of problems, including spreading processes in networks~\cite{taylor2015topological} and the detection of geometric structure in neural activity~\cite{giusti2015clique}. \add{For thorough reviews of current advances in TDA, we refer the interested reader to~\cite{otter2017roadmap,battiston2020networks}.}\\

\indent
Taken together, the advances in network geometry offer a new theoretical framework to gain deep insights into the fundamental principles of complex systems and, more generally, into the physical reality. It is not excluded that the existing results and future advances in this area will lead to fruitful cross-fertilization with other areas of physics.\\

\indent
{\bf\emph{Acknowledgments.}} S.H.\ thanks the Israel Science Foundation, ONR, the BIU Center for Research in Applied Cryptography and Cyber Security, NSF-BSF Grant no.~2019740, and DTRA Grant no.~HDTRA-1-19-1-0016 for financial support.
M.B.\ and M.A.S.\ acknowledge support from: a James S.\ McDonnell Foundation Scholar Award in Complex Systems; the ICREA Academia award, funded by the \textit{Generalitat de Catalunya}; \textit{Agencia estatal de investigaci\'on} project no.\ PID2019-106290GB-C22/AEI/10.13039/501100011033; the Spanish \textit{Ministerio de Ciencia, Innovaci\'on y Universidades} project no.\ FIS2016-76830-C2-2-P (AEI/FEDER, UE); project {\it Mapping Big Data Systems: embedding large complex networks in low-dimensional hidden metric spaces}, \textit{Ayudas Fundaci\'on BBVA a Equipos de Investigaci\'on Cient\'{\i}fica 2017}, and \textit{Generalitat de Catalunya} grant no.~2017SGR1064.
D.K.\ acknowledges support from the NSF grant no.~IIS-1741355, and the ARO grant nos.~W911NF-16-1-0391 and W911NF-17-1-0491.

\onecolumngrid

\end{document}